\documentclass[aps,pra,twocolumn,showpacs]{revtex4-1}

\usepackage{bm,graphicx,textcomp,amssymb,amsmath,dcolumn,color}

\newcommand{\ket}[1]{\left|#1\right>}
\newcommand{\bra}[1]{\left<#1\right|}

\newcommand{\nn}{\nonumber\\}

\newcommand{\bea}{\begin{eqnarray}}
\newcommand{\ea}{\end{eqnarray}}
\newcommand{\eea}{\end{eqnarray}}
\newcommand{\ord}{\,{\cal O}}

\newcommand{\s}[1]{\bar{#1}}
\newcommand{\C}[3]{\hat{C}_{#1,#2}^{#3}}
\newcommand{\Cd}[3]{{\hat{C}_{#1,#2}}^{\dagger #3}}
\newcommand{\N}[3]{{\hat{N}_{#1,#2}}^{#3}}
\newcommand{\crangle}{\rangle^\mathrm{corr}}
\renewcommand{\c}[2]{\hat{c}_{#1,#2}}
\newcommand{\cd}[2]{\hat{c}_{#1,#2}^\dagger}

\begin{document}

\title{Boltzmann relaxation dynamics in the strongly interacting Fermi-Hubbard model}

\author{Friedemann Queisser and Ralf~Sch\"utzhold}

\affiliation{Fakult\"at f\"ur Physik,
Universit\"at Duisburg-Essen, Lotharstra{\ss}e 1, Duisburg 47057, Germany,}

\affiliation{Helmholtz-Zentrum Dresden-Rossendorf, 
Bautzner Landstra{\ss}e 400, 01328 Dresden, Germany,}

\affiliation{Institut f\"ur Theoretische Physik, 
Technische Universit\"at Dresden, 01062 Dresden, Germany.}

\date{\today}

\begin{abstract}
Via the hierarchy of correlations, we study the Mott insulator phase of the 
Fermi-Hubbard model in the limit of strong interactions and derive a 
quantum Boltzmann equation describing its relaxation dynamics. 
In stark contrast to the weakly interacting case, we find that the scattering 
cross sections strongly depend on the momenta of the colliding quasi-particles 
and holes. 
Therefore, the relaxation towards equilibrium crucially depends on the spectrum 
of excitations. 
For example, for particle-hole excitations directly at the minimum of the (direct) 
Mott gap, the scattering cross sections vanish such that these excitations can 
have a very long life-time. 
\end{abstract}

\maketitle

\section{Introduction}

The laws of thermodynamics are very powerful tools in physics with far 
reaching consequences.
However, understanding the microscopic origin of thermal behavior can 
be a very challenging question -- which is also the origin of the famous 
debate between Loschmidt and Boltzmann \cite{L1876,B1877,B1872}.
For classical many-body systems, the relaxation to a thermal 
equilibrium state is typically understood in terms of an 
effective description in the form of a Boltzmann equation \cite{B75}. 
When and where such an effective description is adequate can still be 
a non-trivial question \cite{KWW06,GME11,BCH11,PSSV11,RDO08,Getal12,Ketal16,Netal16}, 
related to the BBGKY hierarchy 
\cite{K46,B46,BG46} 
and chaotic versus integrable behavior. 

For quantum many-body systems, the question of whether and how these 
systems relax to a thermal equilibrium state can be even more involved 
and is being widely discussed in the literature, see, e.g.,  
\cite{EKW09,EKW10,WEFWBH18,PSP18,D91,S94,RDO02,CR10,RS12,R13}. 
For example, the interplay between disorder and interactions can 
have a non-trivial impact on the relaxation dynamics, see, e.g., 
\cite{BAA06,CRFSS11,NH15}.
In the following, we focus on closed quantum lattice systems without 
disorder and dissipation, whose unitary dynamics describes thermalization 
induced by the intrinsic interactions.  
Still, their relaxation and thermalization dynamics can show non-trivial 
features, e.g., it can undergo 
several stages with different time scales, see, e.g., \cite{BBS03,BBW04,KWE11}. 

The thermalization of weakly interacting quantum many-body systems
is typically understood in terms of a quantum version of the Boltzmann 
equation, derived by means of suitable approximation schemes such as 
the Born-Markov approximation \cite{BP02,RK02}.

There are several investigations for one-dimensional systems,
see, e.g., \cite{MWNM98,R09a,R09b,RDYO07,BKL10,KISD11,SVPH14}.
However, due to energy and momentum conservation and potential 
further conservation laws (chaotic versus integrable behavior),
the relaxation dynamics in one dimension displays peculiar features 
and is qualitatively different from that in higher dimensions.  
Thus, these one-dimensional systems are of limited help for 
understanding higher dimensional cases.

\section{The Model}

In order to start filling this gap, we consider the Fermi-Hubbard Hamiltonian 
as a prototypical model for strongly interacting fermions which move on a regular 
lattice given by the hopping matrix $J_{\mu\nu}$ and repel each other via the local 
interaction $U$ 
\begin{align}
\label{Fermi-Hubbard}
\hat{H}
=
-\frac{1}{Z}\sum_{\mu,\nu,s}J_{\mu\nu}\hat c^\dagger_{\mu,s} \hat c_{\nu,s} 
+U\sum_{\mu}\hat n_\mu^\uparrow \hat n_\mu^\downarrow 
\,.
\end{align}
As usual, $\hat c^\dagger_{\mu,s}$ and $\hat c_{\nu,s}$ are the fermionic 
creation and annihilation operators for the lattice sites $\mu$ and $\nu$ 
and the spin $s\in\{\uparrow,\downarrow\}$ with the corresponding number 
operators $\hat n_\mu^s=\hat c^\dagger_{\mu,s} \hat c_{\mu,s}$.
Furthermore, $Z$ denotes the coordination number of the translationally 
invariant lattice, i.e., the number of nearest neighbors. 

In one spatial dimension, the Fermi-Hubbard Hamiltonian~\eqref{Fermi-Hubbard}
is integrable via the Bethe ansatz \cite{LW68} and thus would not display full thermalization
in view of the infinite number of conserved quantities (in addition to the 
impossibility of thermalization via two-body collisions due to energy and momentum
conservation, as mentioned in the Introduction). 
Thus, we focus on higher-dimensional lattices (with large $Z$) in the following. 

In the limit of small interactions $U$, the ground state 
of~\eqref{Fermi-Hubbard} can be described by a Fermi gas 
and is thus metallic for $0<\langle\hat n_\mu^s\rangle<1$. 
For large interactions $U$, however, the structure of the ground 
state changes. 
Assuming half filling $\langle\hat n_\mu^s\rangle=1/2$, 
the repulsion $U$ generates a gap and we obtain the Mott insulator 
state containing one fermion per site 
(plus virtual tunneling corrections), cf.~\cite{H63,IFT98}. 

\section{Hierarchy of Correlations}

For weak interactions $U$, a perturbative expansion in $U$ 
allows us to simplify the equations of motion and to justify the 
Markov approximation (see the Appendix).
For strong interactions $U$, however, this procedure is no longer applicable 
and thus one has to find an alternative approach.

Here, we employ the hierarchy of correlations 
\cite{NS10,QNS12,QKNS14,KNQS14,NQS14,NQS16,QS19} and consider the reduced 
density matrices $\hat\rho_\mu$ for one site and $\hat\rho_{\mu\nu}$ for two 
sites etc. 
After splitting off the correlations via 
$\hat\rho_{\mu\nu}^{\rm corr}=\hat\rho_{\mu\nu}-\hat\rho_\mu\hat\rho_\nu$ 
and so on, we obtain the following hierarchy of evolution equations 
\cite{NS10}
\bea
\label{on-site}
\partial_t\hat\rho_\mu
&=&
f_1(\hat\rho_\nu,\hat\rho_{\mu\nu}^{\rm corr})
\,,
\\
\label{two-site}
\partial_t\hat\rho_{\mu\nu}^{\rm corr}
&=&
f_2(
\hat\rho_\nu,\hat\rho_{\mu\nu}^{\rm corr},\hat\rho_{\mu\nu\sigma}^{\rm corr})
\,,
\\
\label{three-site}
\partial_t\hat\rho_{\mu\nu\sigma}^{\rm corr}
&=&
f_3(
\hat\rho_\nu,\hat\rho_{\mu\nu}^{\rm corr},\hat\rho_{\mu\nu\sigma}^{\rm corr},
\hat\rho_{\mu\nu\sigma\lambda}^{\rm corr})
\,,
\\
\label{four-site}
\partial_t\hat\rho_{\mu\nu\sigma\lambda}^{\rm corr}
&=&
f_4(
\hat\rho_\nu,\hat\rho_{\mu\nu}^{\rm corr},\hat\rho_{\mu\nu\sigma}^{\rm corr},
\hat\rho_{\mu\nu\sigma\lambda}^{\rm corr}, 
\hat\rho_{\mu\nu\sigma\lambda\zeta}^{\rm corr})
\,,
\ea
and in complete analogy for the higher correlators.

In order to truncate this infinite set of recursive equations, we exploit the 
hierarchy of correlations in the formal limit of large coordination numbers 
$Z\to\infty$.
With the arguments outlined in \cite{NS10}, it can be shown that the 
two-site correlations are suppressed via $\hat\rho_{\mu\nu}^{\rm corr}=\ord(1/Z)$
in comparison to the on-site density matrix $\hat\rho_{\mu}=\ord(Z^0)$.
Furthermore, the three-site correlators are suppressed even stronger via 
$\hat\rho_{\mu\nu\sigma}^{\rm corr}=\ord(1/Z^2)$, and so on.
This hierarchy of correlations facilitates the following iterative 
approximation scheme:
To zeroth order in $1/Z$, we may approximate~\eqref{on-site} via 
$\partial_t\hat\rho_\mu\approx f_1(\hat\rho_\nu,0)$ which yields the mean-field 
solution $\hat\rho_\mu^0$.
As the next step, we may insert this solution $\hat\rho_\mu^0$ 
into~\eqref{two-site} and obtain to first order in $1/Z$ the 
approximation 
$\partial_t\hat\rho_{\mu\nu}^{\rm corr}\approx
f_2(\hat\rho_\nu^0,\hat\rho_{\mu\nu}^{\rm corr},0)$
which gives a set of linear and inhomogeneous equations for the 
two-point correlations $\hat\rho_{\mu\nu}^{\rm corr}$.
From this set, we obtain the quasi-particle excitations and their energies. 

Since this set $\partial_t\hat\rho_{\mu\nu}^{\rm corr}\approx
f_2(\hat\rho_\nu^0,\hat\rho_{\mu\nu}^{\rm corr},0)$
of equations is linear in $\hat\rho_{\mu\nu}^{\rm corr}$, 
it does not describe interactions between the quasi-particles and hence
we do not obtain a Boltzmann collision term to first order in $1/Z$.
To this end, we have to go to higher orders in $1/Z$ and study the 
impact of the three-point correlators $\hat\rho_{\mu\nu\sigma}^{\rm corr}$ 
in~\eqref{two-site}.
As one might already expect from the well-known derivation for weak 
interactions (see the Appendix), it is not sufficient to truncate the set of 
equations~\eqref{on-site}-\eqref{four-site} at this stage -- 
we have to include the four-point correlators in order to derive the 
Boltzmann equation (see below). 

Finally, the back-reaction of the quasi-particle fluctuations onto 
the mean field $\hat\rho_\mu$ can be derived by inserting the solution 
for $\hat\rho_{\mu\nu}^{\rm corr}$ back into equation~\eqref{on-site}.

\section{Mott Insulator State}

As explained above, the starting point of the hierarchy is the on-site 
density matrix $\hat\rho_\mu$ or its zeroth-order (mean-field) approximation 
$\hat\rho_\mu^0$.
Assuming a spatially homogeneous state at half filling \cite{footnote}, 
we get the simple solution of equation~\eqref{on-site}
\begin{align} 
\label{double-occupancy}
\hat\rho_\mu
=
\left(\frac{1}{2}-\mathfrak{D}\right)
\left(\ket{\uparrow}\bra{\uparrow}+\ket{\downarrow}\bra{\downarrow}\right)
+\mathfrak{D}
\left(\ket{\uparrow\downarrow}\bra{\uparrow\downarrow}+\ket{0}\bra{0}\right)
\,,
\end{align}
where $\mathfrak{D}$ denotes the double occupancy and measures the deviation 
from the ideal Mott insulator state for $U\gg J$. 

Now we may insert this solution into Eq.~\eqref{two-site} and study the 
two-point correlations. 
In order to describe the relevant correlators describing the dynamics 
of the quasi-particles (also called doublons) and holes (or holons), 
we introduce the short-hand notation $\hat N_{\mu,s}^X$ 
which is just $\hat n_{\mu,s}$ for $X=1$ but 
$1-\hat n_{\mu,s}$ for $X=0$ (see the Appendix).
Then we may define the uppercase operators via 
\bea
\label{uppercase}
\hat C_{\mu,s}^X=\hat c_{\mu,s} \hat N_{\mu,\bar s}^X
\,,
\ea
where $\bar s$ is the spin index opposite to $s$.
For $X=1$, they correspond to the annihilation of a fermion with 
spin $s$ at the lattice site $\mu$ when there is another fermion 
with opposite spin $\bar s$ at that site. 
Thus, this case $X=1$ corresponds to a quasi-particle (doublon) excitation. 
In analogy, the case $X=0$ corresponds to the absence of another
fermion with opposite spin $\bar s$ at that site, i.e., a hole (holon) excitation.

In terms of these operators~\eqref{uppercase}, the quasi-particle and hole 
correlators can be written as 
\bea
\label{two-point}
f^{XY}_{\mu\nu,s}
=
\langle(\hat C_{\mu,s}^X)^\dagger\hat C_{\nu,s}^Y\rangle 
=
\int\limits_\mathbf{k} f^{XY}_{\mathbf{k},s}
\exp\{i\mathbf{k}\cdot\Delta\mathbf{r}_{\mu\nu}
\}
\,,
\ea
where $\Delta\mathbf{r}_{\mu\nu}=\mathbf{r}_\mu-\mathbf{r}_\nu$
denotes the difference between the positions
$\mathbf{r}_\mu$ and $\mathbf{r}_\nu$ 
of the lattice sites $\mu$ and $\nu$.  
Here, we have assumed spatial homogeneity. 
In principle, one could also consider inhomogeneous 
excitations, where these functions which enter the 
Boltzmann equation would acquire an additional position 
coordinate, i.e., $f^{XY}(\mathbf{k},\mathbf{r},s)$  
instead of $f^{XY}(\mathbf{k},s)$. 
Then, the Boltzmann equation would also contain terms 
$\partial f^{XY}(\mathbf{k},\mathbf{r},s)/\partial\mathbf{r}$ 
describing the propagation of the excitations.
However, here we are mainly interested in the collision
terms in the Boltzmann equation and hence we assume spatial 
homogeneity for simplicity. 

\section{Dispersion Relation}

In terms of the $f^{XY}_{\mathbf{k},s}$, 
the evolution equation for the two-point correlators~\eqref{two-point}
obtained from Eq.~\eqref{two-site} reads 
\begin{align}
\label{twopoint_fourier}
i\partial_t f^{XY}_{\mathbf{k},s}
&=
U(Y-X)f^{XY}_{\mathbf{k},s}
+\frac{J_\mathbf{k}}{2}\sum_Z(f^{ZY}_{\mathbf{k},s}-f^{XZ}_{\mathbf{k},s})
\nonumber\\
&
+S_{\mathbf{k},s}^{XY}
\,,
\end{align}
where the source term $S_{\mathbf{k},s}^{XY}$ contains the three-point 
correlators and is suppressed as $1/Z^2$.
Apart from this source term, the set of equations~(\ref{twopoint_fourier}) 
is linear and can be can be diagonalized by means of an orthogonal 
$2\times2$ transformation matrix $O_X^a(\mathbf{k})$, see the Appendix.
We denote the transformed (rotated) correlation functions by lowercase 
superscripts via 
$f_{\mathbf{k},s}^{ab}=2\sum_{XY}O_X^a(\mathbf{k})O_Y^b(\mathbf{k})f^{XY}_\mathbf{k}$.
Thus, the set of equations (\ref{twopoint_fourier}) simplifies to
\begin{align}\label{twopointdiag}
i\partial_t f^{ab}_{\mathbf{k},s}
=
(E^b_\mathbf{k}-E^a_\mathbf{k})f^{ab}_{\mathbf{k},s}+2S_{\mathbf{k},s}^{ab}
\,,
\end{align}
with the quasi-particle ($a=+$) and hole ($a=-$) energies~\cite{LPM69}
\begin{align}
\label{energies}
E^{\pm}_\mathbf{k}=\frac{1}{2}\left(U-J_\mathbf{k}\pm\sqrt{J_\mathbf{k}^2+U^2}\right)\,.
\end{align}
The functions $f^{ab}_{\mathbf{k},s}$ are rapidly oscillating for $a\neq b$
but slowly varying for $a=b$ because of $S_{\mathbf{k},s}^{XY}=\ord(1/Z^2)$.  
Thus, the $1/Z$-expansion (hierarchy of correlations) employed here 
naturally provides a separation of time scales: 
We have rapidly varying quantities whose rate of change is given by the 
eigen-energies~\eqref{energies} or linear combinations thereof,  
while the rate of change of the slowly varying quantities is suppressed 
with $1/Z$ (or even higher). 
As in the weakly interacting case, this separation of time-scales will
be used to justify the Markov approximation. 

In the (Mott insulating) ground state, these correlation functions 
$f^{XY}_{\mathbf{k},s}$ assume the values 
$f^{01}_{\mathbf{k},s}=f^{10}_{\mathbf{k},s}=J_\mathbf{k}/(4\sqrt{U^2+J_\mathbf{k}^2})$, $f^{00}_{\mathbf{k},s}
=1/4+U/(4\sqrt{U^2+J_\mathbf{k}^2})-\mathfrak{D}$
and $f^{11}_{\mathbf{k},s}=1/4-U/(4\sqrt{U^2+J_\mathbf{k}^2})+\mathfrak{D}$, 
see, e.g., \cite{QS19}. 
Hence any deviation from these values indicates a departure from the 
ground state, i.e., an excitation.
As a result, the correlation functions $f^{ab}_{\mathbf{k},s}$ 
determine the excitations present in our system.
Accordingly, we denote the slowly varying quantities $f^{a=b}_{\mathbf{k},s}$ 
as our quasi-particle distribution functions for ($a=b=+$) with 
$f^+_{\mathbf{k},s}$ and the hole distribution function for ($a=b=-$) with 
$f^-_{\mathbf{k},s}$.

\section{Higher Correlations}

As shown above, the rate of change of $f^{a=b}_{\mathbf{k},s}$ 
is determined by the source term $S_{\mathbf{k},s}^{ab}$ 
containing the three-point correlation functions 
\begin{align} 
\langle\N{\rho}{\bar{s}}{X} (\C{\mu}{s}{Y})^\dagger \C{\nu}{s}{Z}\crangle
&=
\int\limits_{\mathbf{p,q}}
G_{\mathbf{p}\mathbf{q},\bar{s}ss}^{XYZ}
e^{i\mathbf{p}\cdot\Delta\mathbf{r}_{\mu\rho} 
+i\mathbf{q}\cdot\Delta\mathbf{r}_{\nu\rho} 
}
\label{partnumber},
\\
\langle\cd{\rho}{s} \c{\rho}{\s{s}} (\C{\mu}{\s{s}}{X})^\dagger \C{\nu}{s}{Y}\crangle
&=
\int\limits_{\mathbf{p,q}}
I^{XY}_{\mathbf{p}\mathbf{q},\s{s}s}
e^{i\mathbf{p}\cdot\Delta\mathbf{r}_{\mu\rho} 
+i\mathbf{q}\cdot\Delta\mathbf{r}_{\nu\rho} 
}
\label{spinflip}
,
\\
\langle\cd{\rho}{s} \cd{\rho}{\s{s}}\C{\mu}{\s{s}}{X}\C{\nu}{s}{Y}\crangle
&=
\int\limits_{\mathbf{p,q}}
H^{XY}_{\mathbf{p}\mathbf{q},\s{s}s}
e^{i\mathbf{p}\cdot\Delta\mathbf{r}_{\mu\rho} 
+i\mathbf{q}\cdot\Delta\mathbf{r}_{\nu\rho} 
}
\label{doublonholon},
\end{align} 
which are of order $1/Z^2$.
The evolution equations for these correlators (\ref{partnumber})-(\ref{spinflip}) 
can be derived from equation~(\ref{three-site}) and read after the rotation 
with $O_X^a(\mathbf{k})$ 
(see the Appendix)
\begin{align}
i\partial_t G_{\mathbf{p}\mathbf{q},\bar{s}ss}^{Xab}
&=
(E^b_\mathbf{q}-E^a_\mathbf{p})
G_{\mathbf{p}\mathbf{q},\bar{s}ss}^{Xab}
+S_{\mathbf{p}\mathbf{q},\bar{s}ss}^{G,Xab}
\,,
\label{partnumberfourier}
\\
i\partial_t I^{ab}_{\mathbf{p}\mathbf{q},\s{s}s}
&=
(E^b_\mathbf{q}-E^a_\mathbf{p})I^{ab}_{\mathbf{p}\mathbf{q},\s{s}s}
+S^{I,ab}_{\mathbf{p}\mathbf{q},\s{s}s}
\,.
\label{spinflipfourier}
\\
i\partial_t H^{ab}_{\mathbf{p}\mathbf{q},\s{s}s}
&=
(E^a_\mathbf{p}+E^b_\mathbf{q}-U)H^{ab}_{\mathbf{p}\mathbf{q},\s{s}s}
+S^{H,ab}_{\mathbf{p}\mathbf{q},\s{s}s}
\,,
\label{doublonholonfourier}
\end{align}
The source terms $S_{\mathbf{p}\mathbf{q},\bar{s}ss}^{G,Xab}$, 
$S^{I,ab}_{\mathbf{p}\mathbf{q},\s{s}s}$, and 
$S^{H,ab}_{\mathbf{p}\mathbf{q},\s{s}s}$ in the above 
equations~(\ref{partnumberfourier})-(\ref{spinflipfourier}) 
contain various combinations of two-point correlators and 
the four-point correlators which are indispensable 
for the Boltzmann collision terms
\begin{align}
\langle(\C{\alpha}{\s{s}}{X})^\dagger \C{\beta}{\s{s}}{Y}
(\C{\mu}{s}{V})^\dagger \C{\nu}{s}{W}\crangle
=
\int\limits_{\mathbf{p,q,k}}
J_{\mathbf{p}\mathbf{q}\mathbf{k},\bar{s}\bar{s}ss}^{XYVW}\times
\nonumber 
\\
e^{i\mathbf{p}\cdot\Delta\mathbf{r}_{\beta\alpha}
+
i\mathbf{q}\cdot\Delta\mathbf{r}_{\mu\alpha}
+
i\mathbf{k}\cdot\Delta\mathbf{r}_{\nu\alpha}
}\label{4point}\,.
\end{align}
Finally, their evolution equation can be derived from Eq.~(\ref{four-site}).
After a rotation with $O_X^a(\mathbf{k})$, 
we find (see the Appendix)
\begin{align}
\label{4pointfourier}
i\partial_t J_{\mathbf{p}\mathbf{q}\mathbf{k},\bar{s}\bar{s}ss}^{abcd}
&=
(-E^a_\mathbf{k+q+p}+E^b_\mathbf{k}-E^c_\mathbf{q}+E^d_\mathbf{k})
J_{\mathbf{p}\mathbf{q}\mathbf{k},\bar{s}\bar{s}ss}^{abcd}
\nonumber
\\
&+S_{\mathbf{p}\mathbf{q}\mathbf{k},\bar{s}\bar{s}ss}^{abcd}
\end{align}
where the source term $S_{\mathbf{p}\mathbf{q}\mathbf{k},\bar{s}\bar{s}ss}^{abcd}$
contains three-point and two-point correlations as well as terms of higher order 
in $1/Z$, such as the five-point correlator, which we neglect. 

\section{Markov Approximation}

In order to arrive at a time-local Boltzmann equation,
the differential equations (\ref{partnumberfourier})-(\ref{spinflipfourier}) 
and (\ref{4pointfourier}) are integrated within the Markov approximation.
All these equations are of the general form $i\partial_tC =\Omega\, C+S$ 
and thus have formally the solution
\begin{align}\label{sol}
C(t)=-i\int\limits_{-\infty}^t dt' S(t')e^{-i\Omega (t-t')}\,.
\end{align}
The source terms $S$ containing the distribution functions are slowly varying,
with their rate of change being suppressed by $1/Z$ or even more,  
in comparison with the rapid oscillations $\Omega=\ord(Z^0)$. 
Hence we may approximate $S(t')\approx S(t)$ in the above integral~(\ref{sol}) 
which gives 
\begin{align}\label{solmarkov}
C(t)\approx-\frac{S(t)}{\Omega-i\epsilon}\,,
\end{align}
with the infinitesimal shift $\epsilon>0$ selecting the retarded solution. 
As usual, this Markov approximation effectively neglects memory effects. 
It allows the elimination of all three-point and four-point correlators  
such that finally only the slowly varying distribution functions remain.
After some algebra (see the Appendix) we arrive at
\begin{align}
\partial_t f_{\mathbf{k},s}^d 
&=
-2\pi\int\limits_\mathbf{p,q}
\sum_{a,b,c}
M^{abcd}_{\mathbf{p+q,p,k-q,k},\s{s}\s{s}ss} 
\times\nonumber\\
&\delta
\left(E_\mathbf{p+q}^a-E_\mathbf{p}^b+E_\mathbf{k-q}^c-E_\mathbf{k}^d\right)
\times\nonumber\\
&\big[
f_{\mathbf{k},s}^{d}f_{\mathbf{p},\s{s}}^{b}
\left(1-f_{\mathbf{k-q},s}^{c}\right)\left(1-f_{\mathbf{p+q},\s{s}}^{a}\right)
\nonumber\\
&-
f^c_{\mathbf{k-q},s}f^a_{\mathbf{p+q},\s{s}}(1-f^d_{\mathbf{k},s})(1-f^b_{\mathbf{p},\s{s}})
\big].
\label{Boltzmann-general}
\end{align}
This is the quantum Boltzmann equation and represents our main result.
It has the same general form as in the weakly interacting case.
Let us first discuss the common features. 
The $M^{abcd}_{\mathbf{p+q,p,k-q,k},\s{s}\s{s}ss}$ describe the scattering cross 
sections for the various processes.
For example, $M^{++++}_{\mathbf{p+q,p,k-q,k},\s{s}\s{s}ss}$ corresponds to the 
collision of two quasi-particles with initial momenta $\mathbf{k}$ and $\mathbf{p}$, 
which are scattered to the final momenta $\mathbf{k-q}$ and $\mathbf{p+q}$, 
thus satisfying momentum conservation (with the momentum transfer $\mathbf{q}$). 
Energy conservation is incorporated via the Dirac delta function in the second line 
of Eq.~\eqref{Boltzmann-general}.
The last line of Eq.~\eqref{Boltzmann-general} corresponds to the inverse process, 
which ensures the conservation of probability. 

As another analogy to the weakly interacting case, the structure of the last two lines 
of Eq.~\eqref{Boltzmann-general} reflects the fermionic character of the 
quasi-particles and holes. 
(For bosons, one would have $1+f^d_{\mathbf{k},s}$ instead of $1-f^d_{\mathbf{k},s}$.)
Related to this fermionic nature is the particle-hole duality where the distribution 
function $f^+_{\mathbf{k},s}$ describing quasi-particles is mapped to the distribution 
function $1-f^-_{\mathbf{k},s}$ of the holes. 
Thus, in addition to $2\to2$ processes such as the collision between two quasi-particles 
$M^{++++}_{\mathbf{p+q,p,k-q,k},\s{s}\s{s}ss}$
or two holes 
$M^{----}_{\mathbf{p+q,p,k-q,k},\s{s}\s{s}ss}$
or a quasi-particle with a hole 
$M^{--++}_{\mathbf{p+q,p,k-q,k},\s{s}\s{s}ss}$,
the above equation~\eqref{Boltzmann-general} does in principle also contain 
$1\to3$ processes:
E.g., $M^{+-++}_{\mathbf{p+q,p,k-q,k},\s{s}\s{s}ss}$ corresponds to 
the inelastic scattering of one quasi-particle via the simultaneous 
creation of a new particle-hole pair (or the inverse process). 
However, here we are mainly interested in the strongly interacting limit $U\gg J$,
where such processes are forbidden by energy conservation: 
The initial particle energy 
$E^{+}_\mathbf{k}\approx U-J_\mathbf{k}/2$ is not large enough to 
create a final state with an energy of nearly $2U$.

As the final analogy to the weakly interacting case, we note that only 
quasi-particles (or holes) of opposite spins $s$ and $\bar s$ scatter,
at least to the leading order considered here.
For weak interactions, this is a simple consequence of the structure 
of the on-site interaction term $U\hat n_\mu^\uparrow \hat n_\mu^\downarrow$, 
but for strong interactions, the situation is a bit more complex (see below). 

\section{Strongly Interacting Limit}

As the most crucial difference to the weakly interacting case, the 
scattering cross sections $M^{abcd}_{\mathbf{p+q,p,k-q,k},\s{s}\s{s}ss}$ 
acquire a non-trivial momentum dependence. 
To illustrate this, let us consider the limit of strong interactions $U\gg J$.
In this limit, the Boltzmann equation~\eqref{Boltzmann-general} describing 
collisions of two quasi-particles simplifies to  
\bea
\label{boltzmann}
\partial_t f_{\mathbf{k},s}^+
&\approx& 
-2\pi\int\limits_\mathbf{p,q}
(J_\mathbf{k}+J_\mathbf{p})^2
\delta\left(J_\mathbf{p+q}-J_\mathbf{p}+J_\mathbf{k-q}-J_\mathbf{k}\right)
\nonumber\\
&&
\Big[f^+_{\mathbf{k},s}f^+_{\mathbf{p},\s{s}}
\left(1-f_{\mathbf{k-q},s}^{+}\right)\left(1-f_{\mathbf{p+q},\s{s}}^{+}\right)
\nonumber\\
&&
-
f^+_{\mathbf{k-q},s}f^+_{\mathbf{p+q},\s{s}}(1-f^+_{\mathbf{k},s})(1-f^+_{\mathbf{p},\s{s}})
\Big]
\,.
\ea
For the collision of two holes, the equation has the same form after replacing 
all the $f^+$ with $f^-$.
The equations describing the collision of a quasi-particle and a hole have a 
very similar structure (see the Appendix). 

For weakly interacting systems (see the Appendix), the scattering cross section is 
momentum independent and given by $U^2$. 
Here, we find that the interaction $U$ does not occur in the Boltzmann 
equation~\eqref{boltzmann} at all, where the scattering cross section  
reads $(J_\mathbf{k}+J_\mathbf{p})^2$ and is thus depends on the momenta 
$\mathbf{k}$ and $\mathbf{p}$ of the incoming quasi-particles. 
This difference can be understood in terms of the following simplified and 
intuitive picture:
In the Mott insulator state, all lattice sites are occupied by one fermion and 
thus a quasi-particle roughly corresponds to a doubly occupied lattice site. 
As a consequence, two quasi-particles cannot occur at the same lattice site 
and thus they cannot directly interact via the strong on-site repulsion $U$.
Instead, they can ``feel'' each other via virtual tunneling processes 
(which are Pauli blocked if the neighboring lattice site is also occupied by 
a quasi-particle).  
These virtual tunneling processes explain the scaling with $J^2$ and the 
momentum dependence. 

This momentum dependence can have strong implications for the relaxation dynamics:
If we consider momentum conserving excitation process such as a long-wavelength 
pump laser, the energy cost of creating a particle-hole pair is given by the direct 
gap 
\bea
\Delta E_\mathbf{k}=E^{+}_\mathbf{k}-E^{-}_\mathbf{k}=\sqrt{J_\mathbf{k}^2+U^2}
\,,
\ea
which assumes its minimum value $\Delta E_\mathbf{k}^{\rm min}=U$ at those points 
where $J_\mathbf{k}$ vanishes. 
Now, a weak enough pump laser with a frequency sufficiently below the gap would 
predominantly create excitations near those minimum-energy wave-numbers $\mathbf{k}$ 
where $J_\mathbf{k}=0$.
On the other hand, for these quasi-particle excitations, the scattering cross sections 
$(J_\mathbf{k}+J_\mathbf{p})^2$ in the Boltzmann equation~\eqref{boltzmann}
vanish and thus they would relax very slowly. 
This behavior is also shown by the other channels (such as particle-hole collisions)
in the strongly interacting limit. 

\section{Back-Reaction}

Finally, via inserting the correlation functions back into equation~\eqref{on-site},
we may calculate the back-reaction of the quasi-particle and hole fluctuations onto 
the mean field $\hat\rho_\mu$.
This determines the double occupancy in Eq.~\eqref{double-occupancy} via 
\bea\label{doubleoccup}
i\partial_t \mathfrak{D}=
\sum_s \int_\mathbf{k}J_\mathbf{k}(f^{01}_{\mathbf{k},s}-f^{10}_{\mathbf{k},s})
\,.
\ea
However, this small double occupancy $\mathfrak{D}=\ord(1/Z)$ does not affect our 
leading-order results, such as the scattering cross sections in the Boltzmann 
equation~\eqref{boltzmann}. 

\section{Conclusions and Outlook}

As a prototypical example for strongly interacting quantum many-body system on 
a lattice, we consider the Fermi-Hubbard model~\eqref{Fermi-Hubbard} in the 
Mott insulator state.
Via the hierarchy of correlations, we derive a quantum Boltzmann 
equation~\eqref{Boltzmann-general} describing the relaxation dynamics of the 
quasi-particle (doublon) and hole (holon) excitations. 
As the most crucial difference to the weakly interacting case, we find that 
the scattering cross sections display a strong momentum dependence, 
cf.~Eq.~\eqref{boltzmann}, which has profound consequences for the relaxation 
dynamics. 
In analogy to the weakly interacting case, the Boltzmann equation~\eqref{boltzmann}
facilitates the derivation of an $H$-theorem.

Our method can be generalized to other lattice systems, 
such as the Bose-Hubbard model or spin lattices \cite{KS18,WT11}. 
It can also be used to study higher-order correlators such as the spin modes 
in the Fermi-Hubbard model 
(such as $\langle\hat\sigma_\mu^x\hat\sigma_\nu^x\rangle^{\rm corr}$ with 
$\hat\sigma_\mu^x=\hat c_{\mu,\uparrow}\hat c_{\mu,\downarrow}^\dagger/2+\rm h.c.$), 
which are of bosonic nature.  
Considering the extended Fermi-Hubbard model including long-range Coulomb 
interactions, one would expect that they generate additional scattering cross 
sections in the Boltzmann equation~\eqref{boltzmann} and thus also influence 
the relaxation dynamics. 

\acknowledgments


This work was funded by DFG, grant \# 278162697 (SFB 1242) and 398912239.



\appendix

\section{Boltzmann equations for weakly interacting fermions}

For weakly interacting fermions, the Boltzmann evolution equation 
can be derived via time-dependent perturbation theory.
The Hamiltonian for interacting fermions reads
\begin{align}\label{hamfermions}
\hat{H}=-\frac{1}{Z}\sum_{\mu,\nu,s}J_{\mu\nu}c^\dagger_{\mu,s} \hat c_{\nu,s}+\frac{1}{2Z}\sum_{\mu,\nu,s,s'}V_{\mu\nu}^{ss'}\hat n_{\mu,s} \hat n_{\nu,s'} 
\end{align}
where $s$ and $s'$ are spin indices and $V_{\mu\nu}^{ss'}$
denotes the interaction potential.
In order to apply perturbation theory, we shall transform 
(\ref{hamfermions}) to Fourier space in order to 
diagonalize the kinetic part.
Note that the hierarchical expansion starts from the atomic 
limit and the hopping Hamiltonian introduces the correlation 
between lattice sites, see below.
The Hamiltonian (\ref{hamfermions}) has the Fourier representation
\begin{align}
\hat{H}&=-\sum_{\mathbf{k},s}J_\mathbf{k}\hat{c}_{\mathbf{k},s}^\dagger \hat{c}_{\mathbf{k},s}\nonumber\\
&
+\frac{1}{2N} \sum_{\mathbf{k},\mathbf{q},\mathbf{p}}\sum_{s,s'}V^{ss'}_\mathbf{k}
\hat{c}_{\mathbf{q+k},s}^\dagger\hat{c}_{\mathbf{q},s}\hat{c}_{\mathbf{p-k}s'}^\dagger\hat{c}_{\mathbf{p},s'}
\end{align}
from which one can obtain the equation of motion of the 
fermion distribution function
$n_{\mathbf{k},s}=\langle \hat{c}_{\mathbf{k},s}^\dagger \hat{c}_{\mathbf{k},s}\rangle$, i.e., 
\begin{align}\label{boltzmannmomentum}
i\partial_t n_{\mathbf{k},s}&=\frac{1}{N}\sum_{\mathbf{q},\mathbf{p}}
\sum_{s'} V_\mathbf{q}^{ss'}
\bigg(\langle\hat{c}^\dagger_{\mathbf{k}s,}\hat{c}^\dagger_{\mathbf{p},s'} \hat{c}_{\mathbf{p+q},s'}\hat{c}_{\mathbf{k-q},s}\rangle^\mathrm{corr}\nonumber\\
&-\langle\hat{c}^\dagger_{\mathbf{k-q},s}\hat{c}^\dagger_{\mathbf{p+q},s'} \hat{c}_{\mathbf{p},s'}\hat{c}_{\mathbf{k},s}\rangle^\mathrm{corr}
\bigg)\,.
\end{align}
As can be seen from (\ref{boltzmannmomentum}), 
the dynamics is solely governed by the  correlation functions
\begin{align}\label{4pointmomentum}
&\langle\hat c^\dagger_{\mathbf{p}_1,s}\hat c^\dagger_{\mathbf{p}_2,s'} \hat c_{\mathbf{p}_3,s'}\hat c_{\mathbf{p}_4,s}\rangle^\mathrm{corr}=
\langle\hat c^\dagger_{\mathbf{p}_1,s}\hat c^\dagger_{\mathbf{p}_2,s'} \hat c_{\mathbf{p}_3,s'}\hat c_{\mathbf{p}_4,s}\rangle\nonumber\\
&+
(\delta_{s,s'}\delta_{\mathbf{p}_1,\mathbf{p}_3}\delta_{\mathbf{p}_2,\mathbf{p}_4}
-\delta_{\mathbf{p}_1,\mathbf{p}_4}\delta_{\mathbf{p}_2,\mathbf{p}_3})n_{\mathbf{p}_1,s}n_{\mathbf{p}_2,s'}\,.
 \end{align}
The equation of motion of the correlators (\ref{4pointmomentum})
can be integrated within Markov approximation.
Substituting the resulting expression into (\ref{boltzmannmomentum})
we arrive at
\begin{widetext}
\begin{align}\label{boltzmannweak}
\partial_t n_{\mathbf{k},s}&=-\frac{2\pi}{N^2}\sum_\mathbf{q,p}
\delta(J_\mathbf{k}+J_\mathbf{p}-J_\mathbf{k-q}-J_\mathbf{p-q})
\nonumber\\
&\times\bigg[\sum_{s',s''}V^{ss'}_\mathbf{q}V^{ss''}_\mathbf{q}\left\{n_{\mathbf{k},s}n_{\mathbf{p},s''}(1-n_{\mathbf{k-q},s})(1-n_{\mathbf{p+q},s''})-
n_{\mathbf{k-q},s}n_{\mathbf{p+q},s''}(1-n_{\mathbf{k},s})(1-n_{\mathbf{p},s''})\right\}\nonumber\\
&-V^{ss}_\mathbf{q}V^{ss}_{\mathbf{k-p-q}}
\left\{n_{\mathbf{k},s}n_{\mathbf{p},s}(1-n_{\mathbf{k-q},s})(1-n_{\mathbf{p+q},s})-
n_{\mathbf{k-q},s}n_{\mathbf{p+q},s}(1-n_{\mathbf{k},s})(1-n_{\mathbf{p},s})\right\}\bigg]\,.
\end{align}
\end{widetext}

\section{Boltzmann equations for the strongly interacting Hubbard model}

It is clear that for strongly interacting systems, the derivation 
of the Boltzmann dynamics cannot be based on an expansion in 
powers of the interaction strength between the electrons.
As explained in the paper, we employ therefore a hierarchical 
expansion for large coordination numbers $Z$.

In the following we give a step-by-step derivation of the 
Boltzmann kinetic equation (22).
We consider the simplest possible case and assume that the 
system is always in an unpolarized state at half filling which is metallic for $U\ll J$ and 
insulating for $U\gg J$.
We demand that the initial state has $\sigma_z$-symmetry, such that the 
density matrix commutes with $\sum_{\mu}(\hat{n}_{\mu,\uparrow}-\hat{n}_{\mu,\downarrow})$ 
for all times.

\subsection{Operator equations}

We introduce a compact notation in order to  
make the calculation tractable.
Therefore we define the operators
\begin{align} 
\N{\mu}{s}{0}&=1-\hat{n}_{\mu,s}=1-\N{\mu}{s}{1}\\
\C{\mu}{s}{X}&=\c{\mu}{s}\N{\mu}{\s{s}}{X} 
\end{align}
where $\mu$ denotes the lattice site and $s$ is the spin index.
Using the Heisenberg equations for the Hubbard Hamiltonian~(1), we find
\begin{align}
i\partial_t \Cd{\mu}{s}{X}&=\frac{1}{Z}\sum_{\kappa,Y}J_{\mu\kappa} \Cd{\kappa}{s}{Y} \N{\mu}{\s{s}}{X}-U^{X}\Cd{\mu}{s}{X}\nonumber\\
&+\frac{(-1)^X}{Z}\sum_\kappa J_{\mu\kappa}[\cd{\mu}{s}\c{\mu}{\s{s}}\cd{\kappa}{\s{s}}+\cd{\mu}{s}\cd{\mu}{\s{s}}\c{\kappa}{\s{s}}]
\end{align}
and
\begin{align}
i\partial_t \C{\mu}{s}{X}&=-\frac{1}{Z}\sum_{\kappa,Y}J_{\mu\kappa} \C{\kappa}{s}{Y} \N{\mu}{\s{s}}{X}+U^{X}\C{\mu}{s}{X}\nonumber\\ 
&-\frac{(-1)^X}{Z}\sum_\kappa J_{\mu\kappa}[\c{\kappa}{\s{s}}\cd{\mu}{\s{s}}\c{\mu}{s}+\cd{\kappa}{\s{s}}\c{\mu}{\s{s}}\c{\mu}{s}]
\end{align}
with $U^0=0$ and $U^1=U$.
The operator $\N{\mu}{s}{X}$ evolves according to
\begin{align}
i\partial_t \N{\mu}{s}{X}=\frac{(-1)^X}{Z}\sum_{\kappa,Y,W}J_{\mu\kappa}\left[\Cd{\mu}{s}{Y}\C{\kappa}{s}{W}-\Cd{\kappa}{s}{Y}\C{\mu}{s}{W} \right]\,,
\end{align}
the spin-flip operator satisfies the equation
\begin{align}
i\partial_t (\cd{\mu}{s}\c{\mu}{\s{s}})=-\frac{1}{Z}\sum_{\kappa,Y,W}J_{\mu\kappa}\left[\Cd{\mu}{s}{Y}\C{\kappa}{\s{s}}{W}-\Cd{\kappa}{s}{Y}\C{\mu}{\s{s}}{W} \right]\,,
\end{align}
and the doublon creation (annihilation) operators have the equation of motion
\begin{align}
i\partial_t(\cd{\mu}{s}\cd{\mu}{\s{s}})&=\frac{1}{Z}\sum_{\kappa,Y,W}J_{\mu\kappa}\left[\Cd{\mu}{s}{Y}\Cd{\kappa}{\s{s}}{W}+\Cd{\kappa}{s}{Y}\Cd{\mu}{\s{s}}{W}\right]\nonumber\\
&-U\cd{\mu}{s}\cd{\mu}{\s{s}}
\end{align}
and
\begin{align}
i\partial_t(\c{\mu}{\s{s}}\c{\mu}{s})&=-\frac{1}{Z}\sum_{\kappa,Y,W}J_{\mu\kappa}\left[\C{\kappa}{\s{s}}{W}\C{\mu}{s}{Y}+\C{\mu}{\s{s}}{W}\C{\kappa}{s}{Y}\right]\nonumber\\
&+U\c{\mu}{\s{s}}\c{\mu}{s}\,.
\end{align}
In the following we shall use the above relations to evaluate 
the evolution equations of the hierarchical correlation functions.

\subsection{Double occupancy and two-site correlation functions}

Due to the $\sigma_z$-symmetry, any expectation value which contains 
an odd number of creation operators 
and annihilation operators for a fixed spin index vanishes identically.
This implies, for example,
\begin{align}
\langle \cd{\mu}{s}\c{\mu}{\s{s}} \rangle=0\quad \text{ or }\quad \langle\Cd{\mu}{s}{X} \C{\nu}{\s{s}}{Y}\crangle= 0\,.
\end{align}
The zeroth order equation of the hierarchical expansion (2) 
determines the double-occupancy $\langle \N{\mu}{s}{1}\N{\mu}{\s{s}}{1}\rangle=
\langle \N{\mu}{s}{0}\N{\mu}{\s{s}}{0}\rangle=~\mathfrak{D}$, i.e.
\begin{align}\label{doubleocc}
i\partial_t \mathfrak{D}&=\frac{1}{Z}\sum_{\kappa,s} J_{\mu\kappa}\left[
\langle\Cd{\mu}{s}{0} \C{\kappa}{s}{1}\crangle-\langle\Cd{\kappa}{s}{1} \C{\mu}{s}{0}\crangle\right]\,.
\end{align}
From the first order equation (3) follows the dynamics of the two-point correlation functions 
\begin{widetext}
\begin{align} \label{twosite}
&i\partial_t\langle\Cd{\mu}{s}{X} \C{\nu}{s}{Y}\crangle=(U^Y-U^X)\langle\Cd{\mu}{s}{X} \C{\nu}{s}{Y}\crangle\nn
&+\frac{1}{Z}\sum_{\kappa,W}J_{\mu\kappa}
\left[\langle \N{\mu}{\s{s}}{X}\rangle \langle \Cd{\kappa}{s}{W} \C{\nu}{s}{Y}\crangle 
+ \langle\N{\mu}{\s{s}}{X} \Cd{\kappa}{s}{W} \C{\nu}{s}{Y}\crangle\right]
-\frac{1}{Z}\sum_{\kappa,W}J_{\nu\kappa}
\left[\langle \N{\nu}{\s{s}}{Y}\rangle \langle \Cd{\mu}{s}{X} \C{\kappa}{s}{W}\crangle 
+ \langle\N{\nu}{\s{s}}{Y} \Cd{\mu}{s}{X} \C{\kappa}{s}{W}\crangle\right]\nn
&+(-1)^X\frac{1}{Z}\sum_{\kappa,W}J_{\mu\kappa}\langle[\cd{\mu}{s} \c{\mu}{\s{s}}\Cd{\kappa}{\s{s}}{W}
+\cd{\mu}{s} \cd{\mu}{\s{s}}\C{\kappa}{\s{s}}{W}]\C{\nu}{s}{Y}\crangle
-(-1)^Y\frac{1}{Z}\sum_{\kappa,W}J_{\nu\kappa}\langle \Cd{\mu}{s}{X}[\C{\kappa}{\s{s}}{W}\cd{\nu}{\s{s}} \c{\nu}{s}
+\Cd{\kappa}{\s{s}}{W}\c{\nu}{\s{s}} \c{\nu}{s}]\crangle\nn
&+\frac{J_{\mu\nu}}{Z}\left[\langle \N{\mu}{\s{s}}{X}\rangle\langle\N{\nu}{s}{1}\N{\nu}{\s{s}}{Y}\rangle+\langle \N{\mu}{\s{s}}{X}\N{\nu}{s}{1}\N{\nu}{\s{s}}{Y}\crangle
\right]
-\frac{J_{\mu\nu}}{Z}\left[\langle \N{\nu}{\s{s}}{Y} \rangle \langle\N{\mu}{s}{1}\N{\mu}{\s{s}}{X}\rangle+ \langle \N{\nu}{\s{s}}{Y}\N{\mu}{s}{1}\N{\mu}{\s{s}}{X}\crangle \right]\nn
&+(-1)^X\frac{J_{\mu\nu}}{Z}\sum_W\langle[\cd{\mu}{s} \c{\mu}{\s{s}}\Cd{\nu}{\s{s}}{W}
+\cd{\mu}{s} \cd{\mu}{\s{s}}\C{\nu}{\s{s}}{W}]\C{\nu}{s}{Y}\crangle
-(-1)^Y\frac{J_{\mu\nu}}{Z}\sum_{W}\langle \Cd{\mu}{s}{X}[\C{\mu}{\s{s}}{W}\cd{\nu}{\s{s}} \c{\nu}{s}
+\Cd{\mu}{\s{s}}{W}\c{\nu}{\s{s}} \c{\nu}{s}]\crangle\nn
&-\frac{\delta_{\mu\nu}}{Z}\sum_{\kappa,W}J_{\mu\kappa}\left[\langle \N{\mu}{\s{s}}{X}\rangle
\langle \Cd{\kappa}{s}{W}\C{\mu}{s}{Y}\crangle-\langle \N{\mu}{\s{s}}{Y}\rangle
\langle \Cd{\kappa}{s}{X}\C{\mu}{s}{W}\crangle\right]\,.
\end{align}
\end{widetext}
The evolution equation (\ref{twosite}) involves terms of order $\mathcal{O}(1/Z)$ 
which determine the free dynamics of the quasi-particles.
In this order, each mode evolves independently.
The three-point correlations of order $\mathcal{O}(1/Z^2)$
couple different modes with each other and are crucial in 
the derivation of the Boltzmann dynamics, see below.
In order to represent equation (\ref{twosite}) momentum space, we define the 
Fourier components of the two-point correlation function 
and the various three-point correlation function to be (cf.~equations (\ref{two-point}) and (\ref{partnumber}-\ref{spinflip}))
\begin{align}
\langle\Cd{\mu}{s}{X} \C{\nu}{s}{Y}\crangle&=\frac{1}{N}\sum_\mathbf{k}f_{\mathbf{k},s}^{XY,\mathrm{corr}}e^{i\mathbf{k}\cdot\Delta\mathbf{x}_{\mu\nu}}\,,
\end{align}

\begin{align}\label{partnumberapp}
\langle\N{\mu}{t}{W} \Cd{\kappa}{s}{X} \C{\nu}{s}{Y}&\crangle\\
=&\frac{1}{N^2}\sum_{\mathbf{p}_1,\mathbf{p}_2}
G_{\mathbf{p}_1,\mathbf{p}_2,tss}^{WXY}e^{i\mathbf{p}_1\cdot \Delta\mathbf{x}_{\kappa\mu}}
e^{i\mathbf{p}_2\cdot \Delta\mathbf{x}_{\nu\mu}}\,,\nonumber
\end{align}

\begin{align}\label{spinflipapp}
\langle\cd{\mu}{s} \c{\mu}{\s{s}}\Cd{\kappa}{\s{s}}{X}\C{\nu}{s}{Y}&\crangle\\
=&
\frac{1}{N^2}\sum_{\mathbf{p}_1,\mathbf{p}_2} I^{XY}_{\mathbf{p}_1,\mathbf{p}_2,\s{s}s}
e^{i\mathbf{p}_1\cdot \Delta\mathbf{x}_{\kappa\mu}}
e^{i\mathbf{p}_2\cdot \Delta \mathbf{x}_{\nu\mu}}\,,\nonumber
\end{align}

\begin{align}\label{doublonholonapp}
\langle\cd{\mu}{s} \cd{\mu}{\s{s}}\C{\kappa}{\s{s}}{X}\C{\nu}{s}{Y}&\crangle\\
=&
\frac{1}{N^2}\sum_{\mathbf{p}_1,\mathbf{p}_2} H^{XY}_{\mathbf{p}_1,\mathbf{p}_2,\s{s}s}
e^{i\mathbf{p}_1\cdot \Delta\mathbf{x}_{\kappa\mu}}
e^{i\mathbf{p}_2\cdot \Delta\mathbf{x}_{\nu\mu}}\,.\nonumber
\end{align}

With these definitions we find from equation (\ref{twosite})
(cf.~equation (\ref{twopoint_fourier}))
\begin{align}\label{1Z}
i\partial_t f^{XY,\mathrm{corr}}_{\mathbf{k},s}&=(U^Y-U^X)f^{XY,\mathrm{corr}}_{\mathbf{k},s}\nonumber\\
&+\frac{J_\mathbf{k}}{2}\sum_W(f^{WY,\mathrm{corr}}_{\mathbf{k},s}-f^{XW,\mathrm{corr}}_{\mathbf{k},s})\nonumber\\
&+S_{\mathbf{k},s}^{XY,1/Z}
+S_{\mathbf{k},s}^{XY,1/Z^2}
\end{align}
with a source term determining the free quasi-particle dynamics,
\begin{align}
S_{\mathbf{k},s}^{XY,1/Z}&=\frac{J_\mathbf{k}}{2}\left[(-1)^X-(-1)^Y\right]\left(\mathfrak{D}-\frac{1}{4}\right)\,,
\end{align}
and a source term of order $\mathcal{O}(1/Z^2)$ which contains the three-point correlators,
\begin{align}
S_{\mathbf{k},s}^{XY,1/Z^2}&=\frac{1}{N}\sum_{\mathbf{q},W}J_\mathbf{q}\Big[G^{XWY}_{\mathbf{q},\mathbf{k},\s{s}ss}
-\left(G^{YWX}_{\mathbf{q},\mathbf{k},\s{s}ss}\right)^*\nonumber\\
&+(-1)^X I^{WY}_{\mathbf{q},\mathbf{k},\s{s}s}-(-1)^Y \left(I^{WX}_{\mathbf{q},\mathbf{k},\s{s}s}\right)^*\nn
&+(-1)^X H^{WY}_{\mathbf{q},\mathbf{k},\s{s}s}-(-1)^Y \left(H^{WX}_{\mathbf{q},\mathbf{k},\s{s}s}\right)^*\Big]+...\label{source1Z2}\,.
\end{align}
We omitted in equation (\ref{source1Z2}) the terms which do not 
contribute to the Boltzmann dynamics in leading order.

It is useful to employ a two-dimensional orthogonal transformation 
which transforms from the $X-Y$
basis to the particle-hole basis.
A general tensor transforms as 
\begin{align}\label{transformation}
T_{\mathbf{p},\mathbf{q},...}^{ab...}=\sum_{X,Y,...}O^a_X(\mathbf{p})O^b_Y(\mathbf{q})\cdot\cdot\cdot
T_{\mathbf{p},\mathbf{q},...}^{XY...}\,.
\end{align}

The orthogonal matrix $O^a_X(\mathbf{k}) $ satisfies the eigenvalue equation
\begin{align}\label{eigenvalue}
\frac{J_\mathbf{k}}{2}\sum_X O^a_X\left(\mathbf{k})=(-E^a_\mathbf{k}+U^Y\right)O^a_Y(\mathbf{k}) \text{ for } Y=0,1 
\end{align}
and has the explicit form
\begin{align}\label{rotation}
O^a_X(\mathbf{k}) =
\begin{pmatrix}
\cos(\phi_\mathbf{k}) & \sin(\phi_\mathbf{k}) \\
-\sin(\phi_\mathbf{k}) & \cos(\phi_\mathbf{k})
\end{pmatrix}
\end{align}
with
\begin{align}
\cos{\phi_\mathbf{k}}=\frac{1}{\sqrt{2}}\left(1+\frac{U}{\sqrt{J_\mathbf{k}^2+U^2}}\right)^{1/2} 
\end{align}
and
\begin{align}
\sin{\phi_\mathbf{k}}=\frac{J_\mathbf{k}}{\sqrt{2}|J_\mathbf{k}|}\left(1-\frac{U}{\sqrt{J_\mathbf{k}^2+U^2}}\right)^{1/2}\,.
\end{align}
The excitation energies of quasi-particles and holes are (cf.~equation (\ref{energies}))
\begin{align}\label{eigenenergies}
E^-_\mathbf{k}&=\frac{1}{2}\left(U-J_\mathbf{k}-\sqrt{J_\mathbf{k}^2+U^2}\right)\\
E^+_\mathbf{k}&=\frac{1}{2}\left(U-J_\mathbf{k}+\sqrt{J_\mathbf{k}^2+U^2}\right)\,.
\end{align}
With the transformation (\ref{transformation}) we can rewrite the equations (\ref{1Z}) 
as (cf.~equation (\ref{twopointdiag}))
\begin{align}\label{eigen1Z}
i\partial_t f^{ab,\mathrm{corr}}_{\mathbf{k},s}=(-E^a_\mathbf{k}+E^b_\mathbf{k})f^{ab,\mathrm{corr}}_{\mathbf{k},s}+ 
S_{\mathbf{k},s}^{ab,1/Z}
+S_{\mathbf{k},s}^{ab,1/Z^2}\,.
\end{align}
After the rotation into the particle-hole basis 
we can separate the slow degrees of freedom
($a=b$) from the fast degrees of freedom ($a\neq b$)
which are changing on a time-scale $\sim1/U$.
Within Markov approximation (cf.~equations (\ref{sol}) and (\ref{solmarkov})) we find

\begin{align}
f^{ab,\mathrm{corr}}_{\mathbf{k},s}=\frac{S_{\mathbf{k},s}^{ab,1/Z}}{E^a_\mathbf{k}-E^b_\mathbf{k}}
+\mathcal{O}(1/Z^2)\quad\text{ for }\quad a\neq b\,.
\end{align}
The slow dynamics is then determined by the evolution of the diagonal elements,
\begin{align}\label{eom1}
i\partial_t f^{aa,\mathrm{corr}}_{\mathbf{k},s}=S_{\mathbf{k},s}^{aa,1/Z^2}\,, 
\end{align}
since the $1/Z$-contributions of the source term in (\ref{eigen1Z}) are vanishing for $a=b$.

The \textit{correlation functions} and the quasi-particle- and hole-\textit{distribution 
functions} (which contains also the on-site contribution of order $\mathcal{O}(1)$) are related by the algebraic relation
\begin{align}\label{distfun}
f^{a}_{\mathbf{k},s}=\frac{1}{2}+\left(\frac{1}{2}-2\,\mathfrak{D}\right)\sum_X(-1)^X  
O^a_X(\mathbf{k})O^a_X(\mathbf{k})+2f^{aa,\mathrm{corr}}_{\mathbf{k},s}\,.
\end{align}
The time-evolution for a negligible change of the 
double occupancy, $\partial_t \mathfrak{D}\approx 0$, is then given by
\begin{align}\label{distfuneom}
i\partial_t f^{a}_{\mathbf{k},s}=2S_{\mathbf{k},s}^{aa,1/Z^2}=
2 \sum_{XY}O^a_X(\mathbf{k})O^a_Y(\mathbf{k})S_{\mathbf{k},s}^{XY,1/Z^2}\,.
\end{align}

The hierarchical method relies on a separation of expectation values into 
correlated and uncorrelated parts.
Since we want to express our final result in terms of quasi-particle and 
hole distribution functions, we need the inversion of the relation
(\ref{distfun}).
It can be checked that up to first order $\mathcal{O}(1/Z)$ we have
\begin{align}
f^{XY,\mathrm{corr}}_{\mathbf{k},s}&=-\frac{1}{4}\delta^{XY}-\delta^{XY}(-1)^X\left(\frac{1}{4}-\mathfrak{D}\right)\nonumber\\
&+\frac{1}{2}\sum_a O^a_X(\mathbf{k}) O^a_Y(\mathbf{k})
f^a_{\mathbf{k},s}+\mathcal{O}(1/Z^2)\,.
\end{align}

\subsection{Boltzmann part of the three-point correlation functions}

The second order of the hierarchical expansion (cf.~equation~(\ref{three-site})) determines the 
evolution of the three-point correlation functions (\ref{partnumberapp}), (\ref{spinflipapp}) 
and (\ref{doublonholonapp}).
Since we are primarily interested in correlations among four lattice sites, 
we shall omit here the explicit form of the source terms which contain only 
two- or three-point correlation functions.
Some of the equations below end therefore with ``...''.

The three-point correlations (\ref{partnumberapp})
are the source terms for \textit{particle-number correlations}.
For them we find  
\begin{align}
i\partial_t \langle \N{\mu}{\s{s}}{W} \Cd{\kappa}{s}{X}&\C{\nu}{s}{Y}\crangle=(U^Y-U^X)\langle \N{\mu}{\s{s}}{W} \Cd{\kappa}{s}{X}\C{\nu}{s}{Y}\crangle\nn
&+\frac{1}{Z}\sum_{\lambda,V}J_{\kappa\lambda}\langle \N{\kappa}{\s{s}}{X}\rangle
\langle \N{\mu}{\s{s}}{W}\Cd{\lambda}{s}{V}\C{\nu}{s}{Y}\crangle\nn
&-\frac{1}{Z}\sum_{\lambda,V}J_{\nu\lambda}\langle \N{\nu}{\s{s}}{Y}\rangle
\langle \N{\mu}{\s{s}}{W}\Cd{\kappa}{s}{X}\C{\lambda}{s}{V}\crangle\nn
&+S_{\mu\kappa\nu,\s{s}ss}^{G,WXY,1/Z^2}+S_{\mu\kappa\nu,\s{s}ss}^{G,WXY,1/Z^3}
\end{align}
with
\begin{align}
&S_{\mu\kappa\nu,\s{s}ss}^{G,WXY,1/Z^3}=\\
&\frac{(-1)^W}{Z}\sum_{\lambda,U,V}J_{\lambda\mu}\langle[\Cd{\mu}{\s{s}}{U}\C{\lambda}{\s{s}}{V}-
\Cd{\lambda}{\s{s}}{U}\C{\mu}{\s{s}}{V}]\Cd{\kappa}{s}{X}\C{\nu}{s}{Y}\crangle+...\nonumber
\end{align}
Taking the Fourier transform, switching to the particle-hole basis and integrating 
within Markov approximation gives (cf.~equation (\ref{partnumberfourier}))
\begin{align}
G_{\mathbf{p}_1,\mathbf{p}_2,\s{s}ss}^{Xab,1/Z^3}=&\frac{i}{i(E_{\mathbf{p}_1}^a-E_{\mathbf{p}_2}^b)-\epsilon} 
S_{\mathbf{p}_1,\mathbf{p}_2,\s{s}ss}^{G,Xab,1/Z^3}+...\nn
=&(-1)^X\frac{1}{N}\sum_\mathbf{q}\sum_{X,Y,U,V}
\frac{i[J_\mathbf{q}-J_{\mathbf{p}_1+\mathbf{p}_2+\mathbf{q}}]}{i(E_{\mathbf{p}_1}^a-E_{\mathbf{p}_2}^b)-\epsilon}
\nonumber\\
&\times O^a_X(\mathbf{p}_1)O^b_Y(\mathbf{p}_2)
J^{UVXY}_{\mathbf{q},\mathbf{p}_1,\mathbf{p}_2,\s{s}\s{s}ss}+...\,,\label{partnumber4point}
\end{align}
where we introduced the Fourier components of the four-point correlations (cf.~equation (\ref{4point})),
\begin{align}
&\langle \Cd{\lambda}{\s{s}}{U}\C{\mu}{\s{s}}{V}\Cd{\kappa}{s}{X}\C{\nu}{s}{Y}\crangle=
\nonumber\\
&
\frac{1}{N^3}\sum_{\mathbf{q}_1,\mathbf{q}_2,\mathbf{q}_3}J^{UVXY}_{\mathbf{q}_1,\mathbf{q}_2,\mathbf{q}_3,\s{s}\s{s}ss }
e^{i\mathbf{q}_1\cdot \Delta\mathbf{x}_{\mu\lambda}}e^{i\mathbf{q}_2\cdot\Delta\mathbf{x}_{\kappa\lambda}}
e^{i\mathbf{q}_3\cdot\Delta\mathbf{x}_{\nu\lambda}}\,.
\end{align}
The correlation functions (\ref{spinflipapp}) are the source of \textit{spin-flip correlations} 
and obey the differential equation
\begin{align}\label{odespinflip}
i\partial_t \langle \cd{\mu}{s} \c{\mu}{\s{s}}\Cd{\kappa}{\s{s}}{X}&\C{\nu}{s}{Y}\crangle=
(U^Y-U^X)\langle \cd{\mu}{s} \c{\mu}{\s{s}}\Cd{\kappa}{\s{s}}{X}\C{\nu}{s}{Y}\crangle\nn
+&\frac{1}{Z}\sum_{\lambda,W}J_{\kappa\lambda}\langle \N{\kappa}{s}{X}\rangle
\langle \cd{\mu}{s} \c{\mu}{\s{s}}\Cd{\lambda}{\s{s}}{W}\C{\nu}{s}{Y}\crangle\nn
-&\frac{1}{Z}\sum_{\lambda,W}J_{\nu\lambda}\langle \N{\nu}{\s{s}}{Y}\rangle
\langle \cd{\mu}{s} \c{\mu}{\s{s}}\Cd{\kappa}{\s{s}}{X}\C{\lambda}{s}{W}\crangle\nn
+&S_{\mu\kappa\nu,\s{s}s}^{I,XY,1/Z^2}+S_{\mu\kappa\nu,\s{s}s}^{I,XY,1/Z^3}
\end{align}
with
\begin{align}
&S_{\mu\kappa\nu,\s{s}s}^{I,XY,1/Z^3}=\\
&\frac{1}{Z}\sum_{\lambda,U,V}J_{\lambda\mu}\langle[\Cd{\lambda}{s}{U}\C{\mu}{\s{s}}{V}-
\Cd{\mu}{s}{U}\C{\lambda}{\s{s}}{V}]\Cd{\kappa}{\s{s}}{X}\C{\nu}{s}{Y}\crangle+...\,.\nonumber
\end{align}
Again, after Fourier transformation and switching to the particle-hole 
basis, we find within Markov approximation (cf.~equation (\ref{spinflipfourier}))
\begin{align}
I_{\mathbf{p}_1,\mathbf{p}_2,\s{s}s}^{ab,1/Z^3}=&\frac{i}{i(E_{\mathbf{p}_1}^a-E_{\mathbf{p}_2}^b)-\epsilon} 
S_{\mathbf{p}_1,\mathbf{p}_2,\s{s}s}^{I,ab,1/Z^3}+...\nn
=&\frac{1}{N}\sum_\mathbf{q}\sum_{U,V,X,Y}
\frac{i[J_\mathbf{q}-J_{\mathbf{p}_1+\mathbf{p}_2+\mathbf{q}}]}{i(E_{\mathbf{p}_1}^a-E_{\mathbf{p}_2}^b)-\epsilon}\nonumber\\
&\times O^a_X(\mathbf{p}_1)O^b_Y(\mathbf{p}_2)
J^{UYXV}_{\mathbf{p}_2,\mathbf{p}_1,\mathbf{q},ss\s{s}\s{s}}+...\,.\label{spinflip4point}
\end{align}
Finally, the correlation functions (\ref{doublonholonapp}) generate the 
\textit{doublon-holon correlations} and evolve according to
\begin{align}\label{odedoublonholon}
&\hspace{-1cm}i\partial_t \langle \cd{\mu}{s} \cd{\mu}{\s{s}}\C{\kappa}{\s{s}}{X}\C{\nu}{s}{Y}\crangle
=\nonumber\\
&
(U^X+U^Y-U)\langle \cd{\mu}{s} \cd{\mu}{\s{s}}\C{\kappa}{\s{s}}{X}\C{\nu}{s}{Y}\crangle\nn
&-\frac{1}{Z}\sum_{\lambda,W}J_{\kappa \lambda}\langle \N{\kappa}{s}{X}\rangle
\langle \cd{\mu}{s} \cd{\mu}{\s{s}}\C{\lambda}{\s{s}}{W}\C{\nu}{s}{Y}\crangle\nn
&-\frac{1}{Z}\sum_{\lambda,W}J_{\nu \lambda}\langle \N{\nu}{\s{s}}{Y}\rangle
\langle \cd{\mu}{s} \cd{\mu}{\s{s}}\C{\kappa}{\s{s}}{X}\C{\lambda}{s}{W}\crangle\nn
&+S_{\mu\kappa\nu,\s{s}s}^{H,XY,1/Z^2}+S_{\mu\kappa\nu,\s{s}s}^{H,XY,1/Z^3}
\end{align}
with
\begin{align}
&S_{\mu\kappa\nu,\s{s}s}^{H,XY,1/Z^3}=\\
&\frac{1}{Z}\sum_{\lambda,U,V}J_{\lambda\mu}\langle[\Cd{\lambda}{s}{U}\Cd{\mu}{\s{s}}{V}+
\Cd{\mu}{s}{U}\Cd{\lambda}{\s{s}}{V}]\C{\kappa}{\s{s}}{X}\C{\nu}{s}{Y}\crangle+...\nonumber
\end{align}
which leads to (cf.~equation (\ref{doublonholonfourier}))
\begin{align}
H_{\mathbf{p}_1,\mathbf{p}_2,\s{s}s}^{ab,1/Z^3}&=
\frac{i}{i(-E_{\mathbf{p}_1}^a-E_{\mathbf{p}_2}^b+U)-\epsilon}S_{\mathbf{p}_1,\mathbf{p}_2,\s{s}s}^{H,ab,1/Z^3}+...\nn
&=
\frac{1}{N}\sum_\mathbf{q}\sum_{X,Y,U,V}
\frac{i[J_\mathbf{q}+J_{\mathbf{p}_1+\mathbf{p}_2+\mathbf{q}}]}{i(-E_{\mathbf{p}_1}^a-E_{\mathbf{p}_2}^b+U)-\epsilon}\nonumber\\
&\times O^a_X(\mathbf{p}_1)O^b_Y(\mathbf{p}_2)
J^{UYVX}_{\mathbf{p}_2,\mathbf{q},\mathbf{p}_1,ss\s{s}\s{s}}+...\,.\label{doublonholon4point}
\end{align}
All these three-point correlators determine the evolution 
of the particle- and hole-distribution functions (\ref{distfun}).
From (\ref{distfuneom}) together with (\ref{partnumber4point}), 
(\ref{spinflip4point}) and (\ref{doublonholon4point}) 
we find

\begin{align}\label{boltzmann1}
&i\partial_t f^{d}_{\mathbf{k},s}=\frac{4}{N^2}\sum_{\mathbf{q},\mathbf{p}}\sum_{X,Y}\sum_{a,b,c}J_\mathbf{q}
(-1)^X O^a_W(\mathbf{k}+\mathbf{q}+\mathbf{p})O^d_X(\mathbf{k})\nn
&\times O^c_Y(\mathbf{q})\Bigg\{
\frac{i[-E_{\mathbf{k}+\mathbf{q}+\mathbf{p}}^a-E_\mathbf{p}^b+U]}
{i(-E^c_\mathbf{q}-E^d_\mathbf{k}+U)-\epsilon}
O^b_{\bar{W}}(\mathbf{p})
J^{adbc}_{\mathbf{k},\mathbf{p},\mathbf{q},ss\s{s}\s{s}}\nonumber\\
&+
\frac{i[E_{\mathbf{k}+\mathbf{q}+\mathbf{p}}^a-E_\mathbf{p}^b]}
{i(E^c_\mathbf{q}-E^d_\mathbf{k})-\epsilon}
O^b_W(\mathbf{p})
\left[J^{abcd}_{\mathbf{p},\mathbf{q},\mathbf{k},\s{s}\s{s}ss}
+J^{adcb}_{\mathbf{k},\mathbf{q},\mathbf{p},ss\s{s}\s{s}}\right]\Bigg\}\nonumber\\
&-c.c.+...\,.
\end{align}

\begin{widetext}

\subsection{Three-point correlation functions up to $1/Z^2$}

In the previous section we omitted the $1/Z^2$-contribution of the 
three-point correlation functions since we focused onto the Boltzmann-part 
which is of order $1/Z^3$.
As will be shown below, the computation of the Fourier components 
$J^{abcd}_{\mathbf{q}_1,\mathbf{q}_2,\mathbf{q}_3,\s{s}\s{s}ss}$
up to order $1/Z^3$ requires the knowledge of $G^{YXab,1/Z^2}_{\mathbf{p}_1,\mathbf{p}_1,\s{s}sss}$,
$I_{\mathbf{p}_1,\mathbf{p}_2,\s{s}s}^{ab,1/Z^2}$ and $H_{\mathbf{p}_1,\mathbf{p}_2,\s{s}s}^{ab,1/Z^2}$.

\subsubsection{Three-point correlators $G^{YXab,1/Z^2}_{\mathbf{p}_1,\mathbf{p}_1,\s{s}sss}$}

We begin with the differential equation for the three-point correlations %
\begin{align}
i\partial_t \langle \N{\mu}{\s{s}}{U} \N{\mu}{s}{V}\Cd{\kappa}{s}{X}\C{\nu}{s}{Y}\crangle
&=\frac{1}{Z}\sum_{\lambda,W}J_{\lambda\kappa}\langle\N{\kappa}{\s{s}}{X}\rangle
\langle \N{\mu}{\s{s}}{U}\N{\mu}{s}{V}\Cd{\lambda}{s}{W}\C{\nu}{s}{Y}\crangle-\frac{1}{Z}\sum_{\lambda,W}J_{\lambda\nu}\langle\N{\nu}{\s{s}}{Y}\rangle
\langle \N{\mu}{\s{s}}{U}\N{\mu}{s}{V}\Cd{\kappa}{s}{X}\C{\lambda}{s}{W}\crangle\nonumber\\
&+(U^Y-U^X)\langle \N{\mu}{\s{s}}{U} \N{\mu}{s}{V}\Cd{\kappa}{s}{X}\C{\nu}{s}{Y}\crangle+S_{\mu\kappa\nu,\s{s}sss}^{G,UVXY,1/Z^2}&\,.
\end{align}
The source term reads
\begin{align}
S_{\mu\kappa\nu,\s{s}sss}^{G,UVXY,1/Z^2}&=
\frac{(-1)^V}{Z}\sum_{\lambda,W} J_{\lambda\mu}[\langle \Cd{\mu}{s}{U}\C{\nu}{s}{Y}\crangle
\langle \C{\lambda}{s}{W}\Cd{\kappa}{s}{X}\crangle
+\langle \Cd{\kappa}{s}{X}\C{\mu}{s}{U}\crangle\langle \Cd{\lambda}{s}{W}\C{\nu}{s}{Y}\crangle]\nonumber\\
&+\frac{(-1)^V}{Z}J_{\kappa\mu}\sum_W \langle \Cd{\mu}{s}{U}\C{\nu}{s}{Y}\crangle \langle \C{\kappa}{s}{W}
\Cd{\kappa}{s}{X}\rangle
+\frac{(-1)^V}{Z}J_{\mu\nu}\sum_W \langle \Cd{\kappa}{s}{X}\C{\mu}{s}{U}\crangle \langle \Cd{\nu}{s}{W}
\C{\nu}{s}{Y}\rangle\nonumber\\
&+\frac{1}{Z}\sum_{\lambda,W}J_{\lambda\kappa}\langle \N{\mu}{\bar{s}}{U}\N{\mu}{s}{V}\N{\kappa}{\bar{s}}{X}\crangle
\langle \Cd{\lambda}{s}{W}\C{\nu}{s}{Y}\crangle\nonumber\\
&+\frac{J_{\mu\kappa}}{Z}\langle \N{\kappa}{\bar{s}}{X}\rangle \langle \N{\mu}{s}{V}\Cd{\mu}{s}{U}
\C{\nu}{s}{Y}\crangle
-\frac{J_{\mu\kappa}}{Z}\sum_W \langle \N{\mu}{\bar{s}}{U}\N{\mu}{s}{V}\rangle \langle \N{\kappa}{\bar{s}}{X}\rangle
\langle \Cd{\mu}{s}{W}\C{\nu}{s}{Y}\crangle\nonumber\\
&+\frac{J_{\nu\alpha}}{Z}\sum_W[\langle \N{\mu}{\bar{s}}{U}\N{\mu}{s}{V}\N{\kappa}{\bar{s}}{X}\crangle
\langle \Cd{\nu}{s}{W}\C{\nu}{s}{Y}\rangle
+\langle \N{\mu}{\bar{s}}{U}\N{\mu}{s}{V}\Cd{\nu}{s}{W}\C{\nu}{s}{Y}\crangle \langle \N{\kappa}{\bar{s}}{X}\rangle]\nonumber\\
&-\frac{1}{Z}\sum_{\lambda,W}J_{\lambda\nu}\langle \N{\mu}{\bar{s}}{U}\N{\mu}{s}{V}\N{\nu}{\bar{s}}{Y}\crangle
\langle \Cd{\kappa}{s}{X}\C{\lambda}{s}{W}\crangle\nonumber\\
&-\frac{J_{\mu\nu}}{Z}\langle \N{\nu}{\bar{s}}{Y}\rangle\langle \N{\mu}{s}{V}\Cd{\kappa}{s}{X}
\C{\mu}{s}{U}\crangle+\frac{J_{\mu\nu}}{Z}\sum_W \langle \N{\mu}{\bar{s}}{U}\N{\mu}{s}{V}\rangle
\langle \N{\nu}{\bar{s}}{Y}\rangle\langle \Cd{\kappa}{s}{X}\C{\mu}{s}{W}\crangle\nonumber\\
&-\frac{J_{\kappa\nu}}{Z}\sum_W[\langle \N{\mu}{\bar{s}}{U}\N{\mu}{s}{V}\N{\nu}{\bar{s}}{Y}\crangle
\langle \Cd{\kappa}{s}{X}\C{\kappa}{s}{W}\rangle+
\langle \N{\mu}{\bar{s}}{U}\N{\mu}{s}{V}\Cd{\kappa}{s}{X}\C{\kappa}{s}{W}\crangle \langle \N{\nu}{\bar{s}}{Y}\rangle]\,.
\end{align}

We neglect the particle-number correlations which are of $\mathcal{O}(1/Z^2)$ and transform 
the Fourier coefficients in the particle-hole basis.
We find the symmetric and anti-symmetric combinations
\begin{align}
\sum_Y G^{YXab,1/Z^2}_{\mathbf{p}_1,\mathbf{p}_2,\s{s}sss}=
\frac{i}{i(E_{\mathbf{p}_1}^a-E_{\mathbf{p}_2}^b)-\epsilon} 
\sum_YS^{G,YXab,1/Z^2}_{\mathbf{p}_1,\mathbf{p}_2,\s{s}sss}
\end{align}
and
\begin{align}
\sum_Y (-1)^YG^{YXab,1/Z^2}_{\mathbf{p}_1,\mathbf{p}_2,\s{s}sss}=
\frac{i}{i(E_{\mathbf{p}_1}^a-E_{\mathbf{p}_2}^b)-\epsilon} 
\sum_Y(-1)^YS^{G,YXab,1/Z^2}_{\mathbf{p}_1,\mathbf{p}_2,\s{s}sss}
\end{align}
with
\begin{align}
\sum_YS^{G,YXab,1/Z^2}_{\mathbf{p}_1,\mathbf{p}_2,\s{s}sss}&=
\frac{(-1)^X}{2}\sum_Y [-E^b_{\mathbf{p}_2}+U^Y]O^a_Y(\mathbf{p}_1)O^b_Y(\mathbf{p}_2)
\left[f^a_{\mathbf{p}_1,s}-\frac{1}{2}-(-1)^Y\left(\frac{1}{2}-2\,\mathfrak{D}\right)\right]\left[f^b_{\mathbf{p}_2,s}-\frac{1}{2}\right]\nonumber\\
&-\frac{(-1)^X}{2}\sum_Y [-E^a_{\mathbf{p}_1}+U^Y]O^a_Y(\mathbf{p}_1)O^b_Y(\mathbf{p}_2)
\left[f^b_{\mathbf{p}_2,s}-\frac{1}{2}-(-1)^Y\left(\frac{1}{2}-2\,\mathfrak{D}\right)\right]\left[f^a_{\mathbf{p}_1,s}-\frac{1}{2}\right]
\end{align}
and
\begin{align}
\sum_Y(-1)^YS^{G,YXab,1/Z^2}_{\mathbf{p}_1,\mathbf{p}_2,\s{s}sss}=&
\frac{(-1)^X}{2}\sum_Y [-E_{\mathbf{p}_2}^b+U^Y]O^a_Y(\mathbf{p}_1)O^b_Y(\mathbf{p}_2)
(-1)^Y\left[f^b_{\mathbf{p}_2,s}-\frac{1}{2}-(-1)^Y\left(\frac{1}{2}-2\,\mathfrak{D}\right)-\frac{1}{2}(-1)^X\right]\nonumber\\
&\times\left[f^a_{\mathbf{p}_1,s}-\frac{1}{2}-(-1)^Y\left(\frac{1}{2}-2\,\mathfrak{D}\right)\right]\nonumber\\
-&\frac{(-1)^X}{2}\sum_Y [-E^a_{\mathbf{p}_1}+U^Y]O^a_Y(\mathbf{p}_1)O^b_Y(\mathbf{p}_2)
(-1)^Y\left[f^a_{\mathbf{p}_1,s}-\frac{1}{2}-(-1)^Y\left(\frac{1}{2}-2\,\mathfrak{D}\right)-\frac{1}{2}(-1)^X\right]\nonumber\\
&\times \left[f^b_{\mathbf{p}_2,s}-\frac{1}{2}-(-1)^Y\left(\frac{1}{2}-2\,\mathfrak{D}\right)\right]\,.
\end{align}

\subsubsection{Three-point correlators $I_{\mathbf{p}_1,\mathbf{p}_2,\s{s}s}^{ab,1/Z^2}$}
The inhomogeneity of order $1/Z^2$ in (\ref{odespinflip}) reads
\begin{align}\label{sourcespinflip1Z2}
S_{\mu\kappa\nu,\s{s}s}^{I,XY,1/Z^2}&=\frac{1}{Z}\sum_{\lambda,U,V}J_{\lambda\mu}
[\langle \Cd{\mu}{s}{U}\C{\nu}{s}{Y}\crangle \langle \Cd{\kappa}{\s{s}}{X}\C{\lambda}{\s{s}}{V}\crangle-
\langle \Cd{\lambda}{s}{U}\C{\nu}{s}{Y}\crangle \langle \Cd{\kappa}{\s{s}}{X}\C{\mu}{\s{s}}{V}\crangle]\nn
&-\frac{J_{\kappa\mu}}{Z}\sum_{U,V}\langle \C{\kappa}{\s{s}}{V}\Cd{\kappa}{\s{s}}{X}\rangle
\langle \Cd{\mu}{s}{U}\C{\nu}{s}{Y}\crangle
+\frac{J_{\mu\nu}}{Z}\sum_{U,V}\langle \Cd{\nu}{s}{U}\C{\nu}{s}{Y}\rangle
\langle\C{\mu}{\s{s}}{V}\Cd{\kappa}{\s{s}}{X}\crangle\nn
&+\frac{J_{\kappa\mu}}{Z}\sum_U \langle \N{\kappa}{s}{X}\rangle\langle \cd{\mu}{s}\c{\mu}{\s{s}}\Cd{\mu}{\s{s}}{U}\C{\nu}{s}{Y}\crangle
-\frac{J_{\mu\nu}}{Z}\sum_U \langle \N{\nu}{\s{s}}{Y}\rangle\langle \cd{\mu}{s}\c{\mu}{\s{s}}\Cd{\kappa}{\s{s}}{X}\C{\mu}{s}{U}\crangle\nn
&+\frac{(-1)^X}{Z}\sum_{\lambda}J_{\kappa\lambda}\langle \cd{\mu}{s}\c{\mu}{\s{s}}\cd{\kappa}{\s{s}}\c{\kappa}{s}\crangle
\langle \cd{\lambda}{s}\C{\nu}{s}{Y}\crangle
-\frac{(-1)^Y}{Z}\sum_\lambda J_{\nu\lambda}\langle \cd{\mu}{s}\c{\mu}{\s{s}}\cd{\nu}{\s{s}}\c{\nu}{s}\crangle \langle \Cd{\kappa}{\s{s}}{X}\c{\lambda}{\s{s}}\crangle\nn
&+\frac{J_{\kappa\nu}}{Z}\left[\sum_U\langle \N{\kappa}{s}{X}\rangle \langle \cd{\mu}{s}\c{\mu}{\s{s}}\Cd{\nu}{\s{s}}{U}\C{\nu}{s}{Y}\crangle
+(-1)^X \langle \cd{\nu}{s}\C{\nu}{s}{Y}\rangle\langle \cd{\mu}{s}\c{\mu}{\s{s}}\cd{\kappa}{\s{s}}\c{\kappa}{s}\crangle\right]\nn
&-\frac{J_{\kappa\nu}}{Z}\left[\sum_U\langle \N{\nu}{\s{s}}{Y}\rangle \langle \cd{\mu}{s}\c{\mu}{\s{s}}\Cd{\kappa}{\s{s}}{X}\C{\kappa}{s}{U}\crangle
+(-1)^Y \langle \Cd{\kappa}{\s{s}}{X}\c{\kappa}{\s{s}}\rangle\langle \cd{\mu}{s}\c{\mu}{\s{s}}\cd{\nu}{\s{s}}\c{\nu}{s}\crangle\right]\,.
\end{align}
The last four lines of equation (\ref{sourcespinflip1Z2}) are of order $1/Z^3$ 
(the two-site correlations of the spin-flip 
operators are of order~$1/Z^2$) and will be neglected in the following.
Within this approximation we arrive at
\begin{align}
I_{\mathbf{p}_1,\mathbf{p}_2,\s{s}s}^{ab,1/Z^2}&=\frac{i}{i(E_{\mathbf{p}_1}^a-E_{\mathbf{p}_2}^b)-\epsilon} 
S_{\mathbf{p}_1,\mathbf{p}_2,\s{s}s}^{I,ab,1/Z^2}
\end{align}
with
\begin{align}
S_{\mathbf{p}_1,\mathbf{p}_2,\s{s}s}^{I,ab,1/Z^2}&=\frac{1}{2}\sum_X[-E_{\mathbf{p}_1}^a+U^X]
O^{a}_{X}(\mathbf{p}_1)O^{b}_{X}(\mathbf{p}_2)\bigg[f^b_{\mathbf{p}_2,s}-\frac{1}{2}-(-1)^X\left(\frac{1}{2}-2\,\mathfrak{D}\right)\bigg]
\bigg[f^a_{\mathbf{p}_1,\s{s}}-\frac{1}{2}+(-1)^X\frac{1}{2}\bigg]\nn
&-\frac{1}{2}\sum_X[-E_{\mathbf{p}_2}^b+U^X]
O^a_X(\mathbf{p}_1)O^b_X(\mathbf{p}_2)\bigg[f^a_{\mathbf{p}_1,\s{s}}-\frac{1}{2}-(-1)^X\left(\frac{1}{2}-2\,\mathfrak{D}\right)\bigg]
\bigg[f^b_{\mathbf{p}_2,s}-\frac{1}{2}+(-1)^X\frac{1}{2}\bigg]\,.
\end{align}

\subsubsection{Three-point correlators $H_{\mathbf{p}_1,\mathbf{p}_2,\s{s}s}^{ab,1/Z^2}$}
The term of order $1/Z^2$ which was omitted in equation (\ref{odedoublonholon}) reads
\begin{align}\label{sourcedoublonholon1Z2}
S_{\mu\kappa\nu,\s{s}s}^{H,XY,1/Z^2}&=\frac{1}{Z}\sum_{\lambda,U,V}J_{\lambda\mu}
[\langle \Cd{\lambda}{s}{U}\C{\nu}{s}{Y}\crangle\langle \Cd{\mu}{\s{s}}{V}\C{\kappa}{\s{s}}{X}\crangle
+\langle \Cd{\lambda}{\s{s}}{U}\C{\kappa}{\s{s}}{X}\crangle \langle \Cd{\mu}{s}{V}\C{\nu}{s}{Y}\crangle]\nn
&+\frac{J_{\kappa\mu}}{Z}\sum_{U,V}\langle \Cd{\kappa}{\s{s}}{U}\C{\kappa}{\s{s}}{X}\rangle
\langle \Cd{\mu}{s}{V}\C{\nu}{s}{Y}\crangle
+\frac{J_{\mu\nu}}{Z}\sum_{U,V}\langle \Cd{\nu}{s}{U}\C{\nu}{s}{Y}\rangle
\langle \Cd{\mu}{\s{s}}{V}\C{\kappa}{\s{s}}{X}\crangle\nn
&-\frac{J_{\kappa\mu}}{Z}\sum_U\langle\N{\kappa}{s}{X}\rangle\langle \cd{\mu}{s}\cd{\mu}{\s{s}}\C{\mu}{\s{s}}{U}\C{\nu}{s}{Y}\crangle
-\frac{J_{\mu\nu}}{Z}\sum_U \langle\N{\nu}{\s{s}}{Y}\rangle \langle \cd{\mu}{s}\cd{\mu}{\s{s}}\C{\kappa}{\s{s}}{X}\C{\mu}{s}{U}\crangle\nn
&+\frac{(-1)^X}{Z}\sum_\lambda J_{\kappa\lambda}\langle \cd{\mu}{s}\cd{\mu}{\s{s}}\c{\kappa}{\s{s}}\c{\kappa}{s}\crangle \langle \cd{\lambda}{s}
\C{\nu}{s}{Y}\crangle
+\frac{(-1)^Y}{Z}\sum_\lambda J_{\nu\lambda}\langle \cd{\mu}{s}\cd{\mu}{\s{s}}\c{\nu}{\s{s}}\c{\nu}{s}\crangle \langle \cd{\lambda}{\s{s}}\C{\kappa}{\s{s}}{X}\crangle\nn
&-\frac{J_{\nu\kappa}}{Z}\bigg[\sum_U \langle \N{\kappa}{s}{X}\rangle\langle \cd{\mu}{s}\cd{\mu}{\s{s}}\C{\nu}{\s{s}}{U}\C{\nu}{s}{Y}\crangle
-(-1)^X\langle \cd{\nu}{s}\C{\nu}{s}{Y}\rangle\langle \cd{\mu}{s}\cd{\mu}{\s{s}}\c{\kappa}{\s{s}}\c{\kappa}{s}\crangle\bigg]\nn
&-\frac{J_{\nu\kappa}}{Z}\bigg[\sum_U\langle \N{\nu}{\s{s}}{Y}\rangle\langle \cd{\mu}{s}\cd{\mu}{\s{s}}
\C{\kappa}{\s{s}}{X}\C{\kappa}{s}{U}\crangle+(-1)^Y\langle \C{\kappa}{\s{s}}{X}\cd{\kappa}{\s{s}}\rangle
\langle \cd{\mu}{s}\cd{\mu}{\s{s}}\c{\nu}{\s{s}}\c{\nu}{s}\crangle\bigg]\,.
\end{align}
Again, the last four lines of equation (\ref{sourcedoublonholon1Z2}) are of order $1/Z^3$ and
will be neglected in the following.
After Fourier transform we obtain within the Markov approximation
\begin{align}
H_{\mathbf{p}_1,\mathbf{p}_2,\s{s}s}^{ab,1/Z^2}&=
\frac{i}{i(-E_{\mathbf{p}_1}^a-E_{\mathbf{p}_2}^b+U)-\epsilon}
S_{\mathbf{p}_1,\mathbf{p}_2,\s{s}s}^{H,ab,1/Z^2}
\end{align}
with
\begin{align}
S_{\mathbf{p}_1,\mathbf{p}_2,\s{s}s}^{H,ab,1/Z^2}&=
\frac{1}{2}\sum_X[-E_{\mathbf{p}_1}^a+U^X]O^a_X(\mathbf{p}_1) O^b_{\s{X}}(\mathbf{p}_2)
\left[f^b_{\mathbf{p}_2,s}-\frac{1}{2}+(-1)^X\left(\frac{1}{2}-2\,\mathfrak{D}\right)\right]\left[f^a_{\mathbf{p}_1,\s{s}}-\frac{1}{2}-(-1)^X\frac{1}{2}\right]\nn
&+\frac{1}{2}\sum_m[-E_{\mathbf{p}_2}^b+U^{\s{X}}]O^a_X(\mathbf{p}_1) O^b_{\s{X}}(\mathbf{p}_2)
\left[f^a_{\mathbf{p}_1,\s{s}}-\frac{1}{2}-(-1)^X\left(\frac{1}{2}-2\,\mathfrak{D}\right)\right]\left[f^b_{\mathbf{p}_2,s}-\frac{1}{2}+(-1)^X\frac{1}{2}\right]\,.
\end{align}

\subsection{Four-point correlation functions up to $1/Z^3$}
The differential equation of the four-point correlators originates from the third 
order of the hierarchical expansion (5) and is given by
\begin{align}
i\partial_t \langle \Cd{\lambda}{\bar{s}}{X}\C{\mu}{\bar{s}}{Y} \Cd{\kappa}{s}{U}\C{\nu}{s}{V}\crangle&=
i\partial_t[\langle \Cd{\lambda}{\bar{s}}{X}\C{\mu}{\bar{s}}{Y} \Cd{\kappa}{s}{U}\C{\nu}{s}{V}\crangle-
\langle \Cd{\lambda}{\bar{s}}{X}\C{\mu}{\bar{s}}{Y}\rangle\langle \Cd{\kappa}{s}{U}\C{\nu}{s}{V}\crangle]\nonumber\\
&=\frac{1}{Z}\sum_{\alpha,W} J_{\alpha\lambda}\langle \N{\lambda}{s}{X} \rangle
\langle \Cd{\alpha}{\bar{s}}{W}\C{\mu}{\bar{s}}{Y}\Cd{\kappa}{s}{U}\C{\nu}{s}{V}\crangle-\frac{1}{Z}\sum_{\alpha,W} J_{\alpha\mu}\langle \N{\mu}{s}{Y} \rangle
\langle \Cd{\lambda}{\bar{s}}{X}\C{\alpha}{\bar{s}}{W}\Cd{\kappa}{s}{U}\C{\nu}{s}{V}\crangle\nonumber\\
&+\frac{1}{Z}\sum_{\alpha,W} J_{\alpha\kappa}\langle \N{\kappa}{\bar{s}}{U} \rangle
\langle \Cd{\lambda}{\bar{s}}{X}\C{\mu}{\bar{s}}{Y}\Cd{\alpha}{s}{W}\C{\nu}{s}{V}\crangle-\frac{1}{Z}\sum_{\alpha,W} J_{\alpha\nu}\langle \N{\mu}{\bar{s}}{V} \rangle
\langle \Cd{\lambda}{\bar{s}}{X}\C{\mu}{\bar{s}}{W}\Cd{\kappa}{s}{U}\C{\alpha}{s}{V}\crangle\nonumber\\
&+(-U^X+U^Y-U^U+U^V)\langle \Cd{\lambda}{\bar{s}}{X}\C{\mu}{\bar{s}}{Y} \Cd{\kappa}{s}{U}\C{\nu}{s}{V}\crangle+S^{J,XYUV,1/Z^3}_{\lambda\mu\kappa\nu,\bar{s}\bar{s}ss}
\end{align}

with
\begin{align}
S^{J,XYUV,1/Z^3}_{\lambda\mu\kappa\nu,\bar{s}\bar{s}ss}=& 
\frac{1}{Z}\sum_{\alpha,W}J_{\alpha\lambda}\big[\langle \N{\lambda}{s}{X}\Cd{\kappa}{s}{U}\C{\nu}{s}{V}\crangle
\langle \Cd{\alpha}{\bar{s}}{W}\C{\mu}{\bar{s}}{Y}\crangle-
(-1)^X\langle \cd{\lambda}{\bar{s}}\c{\lambda}{s}\Cd{\kappa}{s}{U}\C{\mu}{\bar{s}}{Y}\crangle
\langle \Cd{\alpha}{s}{W}\C{\nu}{s}{V}\crangle\nonumber\\
&-(-1)^X \langle \cd{\lambda}{s}\cd{\lambda}{\bar{s}}\C{\mu}{\bar{s}}{Y}\C{\nu}{s}{V}\crangle
\langle \Cd{\kappa}{s}{U}\C{\alpha}{s}{W}\crangle\big]\nonumber\\
-&\frac{1}{Z}\sum_{\alpha,W}J_{\alpha\mu}\big[\langle \N{\mu}{s}{Y}\Cd{\kappa}{s}{U}\C{\nu}{s}{V}\crangle
\langle \Cd{\lambda}{\bar{s}}{X}\C{\alpha}{\bar{s}}{W}\crangle-
(-1)^Y\langle \cd{\mu}{s}\c{\mu}{\bar{s}}\Cd{\lambda}{\bar{s}}{X}\C{\nu}{s}{W}\crangle
\langle \Cd{\kappa}{s}{U}\C{\alpha}{s}{W}\crangle\nonumber\\
&-(-1)^Y \langle \c{\mu}{s}\c{\mu}{\bar{s}}\Cd{\lambda}{\bar{s}}{X}\Cd{\kappa}{s}{U}\crangle
\langle \Cd{\alpha}{s}{W}\C{\nu}{s}{V}\crangle\big]\nonumber\\
+&\frac{1}{Z}\sum_{\alpha,W}J_{\alpha\kappa}\big[\langle \N{\kappa}{\bar{s}}{U}\Cd{\lambda}{\bar{s}}{X}\C{\mu}{\bar{s}}{Y}\crangle
\langle \Cd{\alpha}{s}{W}\C{\nu}{s}{V}\crangle-
(-1)^U\langle \cd{\kappa}{s}\c{\kappa}{\bar{s}}\Cd{\lambda}{\bar{s}}{X}\C{\nu}{s}{V}\crangle
\langle \Cd{\alpha}{\bar{s}}{W}\C{\mu}{\bar{s}}{Y}\crangle\nonumber\\
&-(-1)^U \langle \cd{\kappa}{\bar{s}}\cd{\kappa}{s}\C{\nu}{s}{V}\C{\mu}{\bar{s}}{Y}\crangle
\langle \Cd{\lambda}{\bar{s}}{X}\C{\alpha}{\bar{s}}{W}\crangle\big]\nonumber\\
-&\frac{1}{Z}\sum_{\alpha,W}J_{\alpha\nu}\big[\langle \N{\nu}{\bar{s}}{V}\Cd{\lambda}{\bar{s}}{X}\C{\mu}{\bar{s}}{Y}\crangle
\langle \Cd{\kappa}{s}{U}\C{\alpha}{s}{W}\crangle-
(-1)^V\langle \cd{\nu}{\bar{s}}\c{\nu}{s}\Cd{\kappa}{s}{U}\C{\mu}{\bar{s}}{Y}\crangle
\langle \Cd{\lambda}{\bar{s}}{X}\C{\alpha}{\bar{s}}{W}\crangle\nonumber\\
&-(-1)^V \langle \c{\nu}{s}\c{\nu}{\bar{s}}\Cd{\lambda}{\bar{s}}{X}\Cd{\kappa}{s}{U}\crangle
\langle \Cd{\alpha}{s}{W}\C{\mu}{s}{Y}\crangle\big]\nonumber\\
+&\frac{J_{\kappa\lambda}}{Z}\sum_W[-\langle \N{\lambda}{s}{X}\rangle \langle \Cd{\kappa}{\bar{s}}{W}\Cd{\kappa}{s}{U}
\C{\mu}{\bar{s}}{Y}\C{\nu}{s}{V}\crangle + (-1)^X \langle \C{\kappa}{s}{W}\Cd{\kappa}{s}{U}\rangle
\langle\cd{\lambda}{s}\cd{\lambda}{\bar{s}}\C{\mu}{\bar{s}}{Y}\C{\nu}{s}{V}\crangle\nonumber\\
&-\langle \N{\lambda}{s}{X}\rangle \langle \Cd{\kappa}{\bar{s}}{W}\C{\mu}{\bar{s}}{Y}\crangle \langle \Cd{\kappa}{s}{U}\C{\nu}{s}{V}\crangle
-\langle \N{\kappa}{\bar{s}}{U}\rangle \langle \Cd{\lambda}{\bar{s}}{X}\Cd{\lambda}{s}{W}\C{\mu}{\bar{s}}{Y}\C{\nu}{s}{V}\crangle\nonumber\\
&-\langle \N{\kappa}{\bar{s}}{U}\rangle \langle \Cd{\lambda}{\bar{s}}{X}\C{\mu}{\bar{s}}{Y}\crangle \langle \Cd{\lambda}{s}{W}\C{\lambda}{s}{V}\crangle
-(-1)^U \langle \Cd{\lambda}{\bar{s}}{X}\C{\lambda}{\bar{s}}{W}\rangle \langle \cd{\kappa}{s}\cd{\kappa}{\bar{s}}\C{\mu}{\bar{s}}{Y}\C{\nu}{s}{V}\crangle]\nonumber\\
+&\frac{J_{\nu\lambda}}{Z}\sum_W[-\langle \N{\lambda}{s}{X}\rangle \langle \Cd{\nu}{\bar{s}}{W}\Cd{\nu}{s}{V}
\C{\kappa}{s}{U}\C{\mu}{\bar{s}}{Y}\crangle - (-1)^X \langle \Cd{\nu}{s}{W}\C{\nu}{s}{V}\rangle
\langle\cd{\lambda}{\bar{s}}\c{\lambda}{s}\Cd{\kappa}{s}{U}\C{\mu}{\bar{s}}{Y}\crangle\nonumber\\
&-\langle \N{\lambda}{s}{X}\rangle \langle \Cd{\nu}{\bar{s}}{W}\C{\mu}{\bar{s}}{Y}\crangle \langle \Cd{\kappa}{s}{U}\C{\nu}{s}{V}\crangle
+\langle \N{\nu}{\bar{s}}{V}\rangle \langle \Cd{\lambda}{\bar{s}}{X}\C{\lambda}{s}{W}\Cd{\kappa}{s}{U}\C{\mu}{\bar{s}}{Y}\crangle\nonumber\\
&+\langle \N{\nu}{\bar{s}}{V}\rangle \langle \Cd{\kappa}{s}{U}\C{\lambda}{s}{W}\crangle \langle \Cd{\lambda}{s}{X}\C{\mu}{\bar{s}}{Y}\crangle
-(-1)^V \langle \Cd{\lambda}{\bar{s}}{X}\C{\lambda}{\bar{s}}{W}\rangle \langle \cd{\nu}{\bar{s}}\c{\nu}{s}\C{\kappa}{s}{U}\C{\mu}{\bar{s}}{Y}\crangle]\nonumber\\
+&\frac{J_{\kappa\mu}}{Z}\sum_W[-\langle \N{\mu}{s}{Y}\rangle \langle \C{\kappa}{\bar{s}}{W}\Cd{\kappa}{s}{U}
\Cd{\lambda}{\bar{s}}{X}\C{\nu}{s}{V}\crangle +\langle \N{\mu}{s}{Y}\rangle \langle \Cd{\lambda}{\bar{s}}{X}\C{\kappa}{\bar{s}}{W}\crangle 
\langle \Cd{\kappa}{s}{U}\C{\nu}{s}{V}\crangle\nonumber\\ 
&-(-1)^Y \langle \cd{\mu}{s}\c{\mu}{\bar{s}}\Cd{\lambda}{\bar{s}}{X}\C{\nu}{s}{V}\crangle \langle \C{\kappa}{s}{W} \Cd{\kappa}{s}{U}\rangle
+\langle \N{\kappa}{\bar{s}}{U}\rangle\langle \C{\mu}{\bar{s}}{Y}\Cd{\mu}{s}{W}\Cd{\lambda}{\bar{s}}{X}\C{\nu}{s}{V}\crangle\nonumber\\
&-\langle \N{\kappa}{\bar{s}}{U}\rangle \langle\Cd{\mu}{s}{W}\C{\nu}{s}{V}\crangle\langle\Cd{\lambda}{\bar{s}}{X}\C{\mu}{\bar{s}}{Y}\crangle
+(-1)^U \langle\C{\mu}{\bar{s}}{Y}\Cd{\mu}{\bar{s}}{W}\rangle \langle \cd{\kappa}{s}\c{\kappa}{\bar{s}}\Cd{\lambda}{\bar{s}}{X}\C{\nu}{s}{V}\crangle]\nonumber\\
+&\frac{J_{\mu\nu}}{Z}\sum_W[-\langle \N{\mu}{s}{Y}\rangle \langle \C{\nu}{\bar{s}}{W}\C{\nu}{s}{V}
\Cd{\kappa}{s}{U}\Cd{\lambda}{\bar{s}}{X}\crangle +\langle \N{\mu}{s}{Y}\rangle \langle \Cd{\lambda}{\bar{s}}{X}\C{\nu}{\bar{s}}{W}\crangle 
\langle \Cd{\kappa}{s}{U}\C{\nu}{s}{V}\crangle\nonumber\\ 
&+(-1)^Y \langle \c{\mu}{s}\c{\mu}{\bar{s}}\Cd{\lambda}{\bar{s}}{X}\Cd{\kappa}{s}{U}\crangle \langle \Cd{\nu}{s}{W} \C{\nu}{s}{V}\rangle
-\langle \N{\nu}{\bar{s}}{V}\rangle\langle \C{\mu}{\bar{s}}{Y}\C{\mu}{s}{W}\Cd{\kappa}{s}{U}\Cd{\lambda}{\bar{s}}{X}\crangle\nonumber\\
&+\langle \N{\nu}{\bar{s}}{V}\rangle \langle\Cd{\lambda}{\bar{s}}{X}\C{\mu}{\bar{s}}{Y}\crangle\langle\Cd{\kappa}{s}{U}\C{\mu}{s}{W}\crangle
-(-1)^V \langle\C{\mu}{\bar{s}}{Y}\Cd{\mu}{\bar{s}}{W}\rangle \langle \c{\nu}{\bar{s}}\c{\nu}{s}\Cd{\kappa}{s}{U}\Cd{\lambda}{\bar{s}}{X}\crangle]\nonumber\\
+&\frac{J_{\mu\lambda}}{Z}\sum_W[\langle \Cd{\mu}{\bar{s}}{W}\C{\mu}{\bar{s}}{Y}\crangle \langle\N{\lambda}{s}{X}\Cd{\kappa}{s}{U}\C{\nu}{s}{V}\crangle
+\langle \N{\lambda}{s}{X}\rangle \langle\Cd{\mu}{\bar{s}}{W}\C{\mu}{\bar{s}}{Y}\Cd{\kappa}{s}{U}\C{\nu}{s}{V}\crangle\nonumber\\
&-\langle \N{\mu}{s}{Y}\rangle \langle \Cd{\lambda}{\bar{s}}{X}\C{\lambda}{\bar{s}}{W}\Cd{\kappa}{s}{U}\C{\nu}{s}{V}\crangle
-\langle \Cd{\lambda}{\bar{s}}{X}\C{\lambda}{\bar{s}}{W}\crangle \langle \N{\mu}{s}{Y}\Cd{\kappa}{s}{U}\C{\nu}{s}{V}\crangle]\nonumber\\
+&\frac{J_{\kappa\nu}}{Z}\sum_W[\langle \Cd{\nu}{s}{W}\C{\nu}{s}{V}\crangle \langle\N{\kappa}{\bar{s}}{U}\Cd{\lambda}{\bar{s}}{X}\C{\mu}{\bar{s}}{Y}\crangle
+\langle \N{\kappa}{\bar{s}}{U}\rangle \langle\Cd{\nu}{s}{W}\C{\nu}{s}{V}\Cd{\lambda}{\bar{s}}{X}\C{\mu}{\bar{s}}{Y}\crangle\nonumber\\
&-\langle \N{\nu}{\bar{s}}{V}\rangle \langle \Cd{\kappa}{s}{U}\C{\kappa}{s}{W}\Cd{\lambda}{\bar{s}}{X}\C{\mu}{\bar{s}}{Y}\crangle
-\langle \Cd{\kappa}{s}{U}\C{\kappa}{s}{W}\crangle \langle \N{\nu}{\bar{s}}{V}\Cd{\lambda}{\bar{s}}{Y}\C{\mu}{\bar{s}}{Y}\crangle]
\end{align}
At half filling we find after the Fourier transform 
in Markov approximation
\begin{align}
J^{abcd,1/Z^3}_{-\mathbf{p}_1-\mathbf{p}_2-\mathbf{p}_3,\mathbf{p}_1,\mathbf{p}_2,\mathbf{p}_3,\bar{s}\bar{s}ss}
=\frac{i S^{J,abcd,1/Z^3}_{-\mathbf{p}_1-\mathbf{p}_2-\mathbf{p}_3,\mathbf{p}_1,\mathbf{p}_2,\mathbf{p}_3,\bar{s}\bar{s}ss}}
{i(E_{\mathbf{p}_1+\mathbf{p}_2+\mathbf{p}_3}^a-E_{\mathbf{p}_1}^b
+E_{\mathbf{p}_2}^c-E_{\mathbf{p}_3}^d)-\epsilon}
\end{align}

with the source term
\begin{align}\label{4point1}
S^{J,abcd,1/Z^3}_{-\mathbf{p}_1-\mathbf{p}_2-\mathbf{p}_3,\mathbf{p}_1,\mathbf{p}_2,\mathbf{p}_3,\bar{s}\bar{s}ss}
=&-i\sum_{X,Y}\frac{O^a_X(\mathbf{p}_1+\mathbf{p}_2+\mathbf{p}_3)
O^b_X(\mathbf{p}_1)S^{G,YXcd,1/Z^2}_{\mathbf{p}_2,\mathbf{p}_3,\bar{s}sss}}
{i(E_{\mathbf{p}_2}^c-E_{\mathbf{p}_3}^d)-\epsilon}
\bigg\{
\frac{(-1)^Y}{2}\left[E_{\mathbf{p}_1}^b-E_{\mathbf{p}_1+\mathbf{p}_2+\mathbf{p}_3}^a\right]\nonumber\\
&
+\left[f^a_{\mathbf{p}_1+\mathbf{p}_2+\mathbf{p}_3,\bar{s}}
-\frac{1}{2}\right]\left[-E_{\mathbf{p}_1+\mathbf{p}_2+\mathbf{p}_3}^a+U^X\right]
-\left[f^b_{\mathbf{p}_1,\bar{s}}-\frac{1}{2}\right]\left[-E_{\mathbf{p}_1}^b+U^X\right]
\bigg\}\nonumber\\
-&i\sum_{X,Y}\frac{O^c_X(\mathbf{p}_2)O^d_X(\mathbf{p}_3)
S_{-\mathbf{p}_1-\mathbf{p}_2-\mathbf{p}_3,\mathbf{p}_1,s\bar{s}\bar{s}\bar{s}}^{G,YXab,1/Z^2}}
{i(E_{\mathbf{p}_1+\mathbf{p}_2+\mathbf{p}_3}^a-E_{\mathbf{p}_1}^b)-\epsilon}
\bigg\{
\frac{(-1)^Y}{2}(E_{\mathbf{p}_3}^d-E_{\mathbf{p}_2}^c)\nonumber\\
&+\left[f^c_{\mathbf{p}_2,s}-\frac{1}{2}\right]\left[-E_{\mathbf{p}_2}^c+U^X\right]-
\left[f^d_{\mathbf{p}_3,s}-\frac{1}{2}\right]\left[-E_{\mathbf{p}_3}^d+U^X\right]\bigg\}\nonumber\\
-&i\sum_X \frac{O^b_X(\mathbf{p}_1) O^c_X(\mathbf{p}_2)
S^{I,ad,1/Z^2}_{-\mathbf{p}_1-\mathbf{p}_2-\mathbf{p}_3,\mathbf{p}_3,\bar{s}s}}
{i(E_{\mathbf{p}_1+\mathbf{p}_2+\mathbf{p}_3}^a-E_{\mathbf{p}_3}^d)-\epsilon}
 \bigg\{\left[(-1)^X f^b_{\mathbf{p}_1,\bar{s}}-\frac{(-1)^X}{2}-\frac{1}{2} \right]
 \left[-E_{\mathbf{p}_1}^b+U^X\right]\nonumber\\
&-\left[(-1)^X f^c_{\mathbf{p}_2,s}-\frac{(-1)^X}{2}-\frac{1}{2}\right]\left[-E^c_{\mathbf{p}_2}+U^X\right]\bigg\}
\nonumber\\
-&i\sum_X \frac{O^a_X(\mathbf{p}_1+\mathbf{p}_2+\mathbf{p}_3) O^d_X(\mathbf{p}_3)
S^{I,cb,1/Z^2}_{\mathbf{p}_2,\mathbf{p}_1,s\bar{s}}}
{i(E_{\mathbf{p}_2}^c-E_{\mathbf{p}_1}^b)-\epsilon}
 \bigg\{\left[(-1)^X f^d_{\mathbf{p}_3,s}-\frac{(-1)^X}{2}-\frac{1}{2} \right]
 \left[-E_{\mathbf{p}_3}^d+U^X\right]\nonumber\\
&-\left[(-1)^X f^a_{\mathbf{p}_1+\mathbf{p}_2+\mathbf{p}_3,\bar{s}}-\frac{(-1)^X}{2}
-\frac{1}{2}\right]\left[-E^a_{\mathbf{p}_1+\mathbf{p}_2+\mathbf{p}_3}+U^X\right]\bigg\}\nonumber\\
-&i\sum_X \frac{O^a_X(\mathbf{p}_1+\mathbf{p}_2+\mathbf{p}_3) O^c_X(\mathbf{p}_2)
S^{H,bd,1/Z^2}_{\mathbf{p}_1,\mathbf{p}_3,\bar{s}s}}
{i(-E_{\mathbf{p}_1}^b-E_{\mathbf{p}_3}^d+U)-\epsilon}
 \bigg\{\left[(-1)^X f^c_{\mathbf{p}_2,s}-\frac{(-1)^X}{2}-\frac{1}{2} \right]
 \left[-E_{\mathbf{p}_2}^c+U^X\right]\nonumber\\
&+\left[(-1)^X f^a_{\mathbf{p}_1+\mathbf{p}_2+\mathbf{p}_3,\bar{s}}-\frac{(-1)^X}{2}
-\frac{1}{2}\right]\left[-E^a_{\mathbf{p}_1+\mathbf{p}_2+\mathbf{p}_3}+U^X\right]\bigg\}\nonumber\\
-&i\sum_X \frac{O^b_X(\mathbf{p}_1) O^d_X(\mathbf{p}_3)
S^{H,ca,1/Z^2}_{\mathbf{p}_2,\mathbf{p}_1+\mathbf{p}_2+\mathbf{p}_3s\bar{s}}}
{i(E_{\mathbf{p}_2}^c+E_{\mathbf{p}_1+\mathbf{p}_2+\mathbf{p}_3}^a-U)-\epsilon}
 \bigg\{\left[(-1)^X f^d_{\mathbf{p}_3,s}-\frac{(-1)^X}{2}-\frac{1}{2} \right]
 \left[-E_{\mathbf{p}_3}^d+U^X\right]\nonumber\\
&+2\left[(-1)^X f^b_{\mathbf{p}_1,\bar{s}}-\frac{(-1)^X}{2}
-\frac{1}{2}\right]\left[-E^b_{\mathbf{p}_1}+U^X\right]\bigg\}\nonumber\\
+&\sum_{X,Y}O^a_X(\mathbf{p}_1+\mathbf{p}_2+\mathbf{p}_3) O^b_X(\mathbf{p}_1)
O^c_Y(\mathbf{p}_2)O^d_Y(\mathbf{p}_3)
\bigg\{f_{\mathbf{p}_2,s}^c+f_{\mathbf{p}_3,s}^d-1-(-1)^Y\left(\frac{1}{2}-2\,\mathfrak{D}\right)\bigg\}\nonumber\\
&\times\bigg\{(-E_{\mathbf{p}_1}^b+U^X)
\left(f^a_{\mathbf{p}_1+\mathbf{p}_2+\mathbf{p}_3,\bar{s}}-\frac{1}{2}-(-1)^X\left(\frac{1}{2}-2\,\mathfrak{D}\right)\right)\nonumber\\
&-(-E_{\mathbf{p}_1+\mathbf{p}_2+\mathbf{p}_3}^a+U^X)
\left(f^b_{\mathbf{p}_1,\bar{s}}-\frac{1}{2}-(-1)^X\left(\frac{1}{2}-2\,\mathfrak{D}\right)\right)
\bigg\}\nonumber\\
+&\sum_{X,Y}O^a_Y(\mathbf{p}_1+\mathbf{p}_2+\mathbf{p}_3) O^b_Y(\mathbf{p}_1)
O^c_X(\mathbf{p}_2)O^d_X(\mathbf{p}_3)
\bigg\{f_{\mathbf{p}_1+\mathbf{p}_2+\mathbf{p}_3,\bar{s}}^a+f_{\mathbf{p}_1,\bar{s}}^b-1-(-1)^X\left(\frac{1}{2}-2\,\mathfrak{D}\right)\bigg\}\nonumber\\
&\times\bigg\{(-E_{\mathbf{p}_3}^d+U^X)
\left(f^c_{\mathbf{p}_2,s}-\frac{1}{2}
-(-1)^X\left(\frac{1}{2}-2\,\mathfrak{D}\right)\right)
\nonumber\\
&-(-E_{\mathbf{p}_2}^c+U^X)
\left(f^d_{\mathbf{p}_3,s}-\frac{1}{2}-
(-1)^X\left(\frac{1}{2}-2\,\mathfrak{D}\right)\right)
\bigg\}\,.
\end{align}

\subsection{Boltzmann equations}

From equations (\ref{boltzmann1}) and (\ref{4point1}) we find
after some tedious algebra the time-evolution of the distribution functions 
$f^d_{\mathbf{k},s}$, (cf.~equation (\ref{Boltzmann-general})),
\begin{align}\label{boltzmannevolution}
\partial_t f^d_{\mathbf{k},s}&=\frac{8\pi}{N^2}\sum_{a,b,c}\sum_{\mathbf{q},\mathbf{p}}\sum_{X,Y,V}(-1)^X
\delta(E_{\mathbf{k}+\mathbf{q}+\mathbf{p}}^a-E_\mathbf{p}^b+E_\mathbf{q}^c-E^d_\mathbf{k})\nonumber\\
&\times\big\{J_{\mathbf{k}+\mathbf{q}+\mathbf{p}}O^a_Y(\mathbf{k}+\mathbf{q}+\mathbf{p})O^b_V(\mathbf{p})O^c_V(\mathbf{q})O^d_X(\mathbf{k})
-J_{\mathbf{p}}O^a_V(\mathbf{k}+\mathbf{q}+\mathbf{p})O^b_Y(\mathbf{p})O^c_V(\mathbf{q})O^d_X(\mathbf{k})\nonumber\\
&+J_{\mathbf{q}}O^a_V(\mathbf{k}+\mathbf{q}+\mathbf{p})O^b_V(\mathbf{p})O^c_Y(\mathbf{q})O^d_X(\mathbf{k})
\big\}\mathcal{A}_{-\mathbf{k}-\mathbf{q}-\mathbf{p},\mathbf{k},\mathbf{q},\mathbf{p},ss\bar{s}\bar{s}}^{adcb}
\end{align}
with
\begin{align}\label{4pcorr}
&\mathcal{A}^{abcd}_{-\mathbf{k}-\mathbf{q}-\mathbf{p},\mathbf{p},\mathbf{q},\mathbf{k}, \bar{s}\bar{s}ss}=\nonumber\\
&-\sum_{X,Y,V}\frac{(-1)^X}{16}\bigg[J_{\mathbf{k}+\mathbf{q}+\mathbf{p}}
O^a_Y(\mathbf{k}+\mathbf{q}+\mathbf{p})\left\{ O^b_X(\mathbf{p}) O^c_V(\mathbf{q}) O^d_V(\mathbf{k})- O^b_Z(\mathbf{p}) O^c_X(\mathbf{q}) O^d_{\bar{V}}(\mathbf{k})
+ O^b_V(\mathbf{p}) O^c_V(\mathbf{q}) O^d_X(\mathbf{k})\right\}\nonumber\\
&+J_\mathbf{p}O^b_Y(\mathbf{p})\left\{O^a_X(\mathbf{k}+\mathbf{q}+\mathbf{p})  O^c_V(\mathbf{q}) O^d_V(\mathbf{k})
-O^a_V(\mathbf{k}+\mathbf{q}+\mathbf{p})  O^c_{\bar{V}}(\mathbf{q}) O^d_X(\mathbf{k})
+O^a_V(\mathbf{k}+\mathbf{q}+\mathbf{p})  O^c_X(\mathbf{q}) O^d_V(\mathbf{k})\right\}\nonumber\\
&+J_\mathbf{q}O^c_Y(\mathbf{q})\left\{O^a_V(\mathbf{k}+\mathbf{q}+\mathbf{p}) O^b_V(\mathbf{p})  O^d_X(\mathbf{k})
-O^a_X(\mathbf{k}+\mathbf{q}+\mathbf{p}) O^b_V(\mathbf{p})  O^d_{\bar{V}}(\mathbf{k})
+O^a_V(\mathbf{k}+\mathbf{q}+\mathbf{p}) O^b_X(\mathbf{p})  O^d_V(\mathbf{k})\right\}\nonumber\\
&+J_\mathbf{k}O^d_Y(\mathbf{k})\left\{O^a_V(\mathbf{k}+\mathbf{q}+\mathbf{p}) O^b_V(\mathbf{p}) O^c_X(\mathbf{q}) 
-O^a_V(\mathbf{k}+\mathbf{q}+\mathbf{p}) O^b_X(\mathbf{p}) O^c_{\bar{V}}(\mathbf{q}) 
+O^a_X(\mathbf{k}+\mathbf{q}+\mathbf{p}) O^b_V(\mathbf{p}) O^c_V(\mathbf{q}) \right\}\bigg]\nonumber\\
&\times \left[f^b_{\mathbf{p},\bar{s}} f^d_{\mathbf{k},s}\left(1-f^a_{\mathbf{k}+\mathbf{q}+\mathbf{p},\bar{s}}\right)
\left(1-f^c_{{\mathbf{q}},s}\right)-f^a_{\mathbf{k}+\mathbf{q}+\mathbf{p},\bar{s}}
f^c_{\mathbf{q},s}\left(1-f^b_{\mathbf{p},\bar{s}}\right)\left(1-f^d_{\mathbf{k},s}\right)
\right]\,.
\end{align}
The time-evolution of the double-occupancy is determined by (\ref{doubleocc}).
Within the Markov approximation, its Boltzmann-time-evolution reads (cf.~equation (\ref{doubleoccup}))
\begin{align}
\partial_t \mathfrak{D}&=-\frac{4\pi}{N^3}\sum_s\sum_{a,b,c,d}\sum_{\mathbf{k},\mathbf{q},\mathbf{p}}\frac{J_\mathbf{k}}{\sqrt{J_\mathbf{k}^2+U^2}}\sum_{X,Y,V}(-1)^X
\delta(E_{\mathbf{k}+\mathbf{q}+\mathbf{p}}^a-E_\mathbf{p}^b+E_\mathbf{q}^c-E^d_\mathbf{k})\nonumber\\
&\times\big\{J_{\mathbf{k}+\mathbf{q}+\mathbf{p}}O^a_Y(\mathbf{k}+\mathbf{q}+\mathbf{p})O^b_V(\mathbf{p})O^c_V(\mathbf{q})O^{\s{d}}_X(\mathbf{k})
-J_{\mathbf{p}}O^a_V(\mathbf{k}+\mathbf{q}+\mathbf{p})O^b_Y(\mathbf{p})O^c_{\s{V}}(\mathbf{q})O^{\s{d}}_X(\mathbf{k})\nonumber\\
&+J_{\mathbf{q}}O^a_V(\mathbf{k}+\mathbf{q}+\mathbf{p})O^b_Z(\mathbf{p})O^c_Y(\mathbf{q})O^{\s{d}}_X(\mathbf{k})
\big\}\mathcal{A}_{-\mathbf{k}-\mathbf{q}-\mathbf{p},\mathbf{k},\mathbf{q},\mathbf{p},ss\bar{s}\bar{s}}^{adcb}
\end{align}
which is of order $\mathcal{O}(1/Z^4)$ and becomes negligible for $J\ll U$.

%
%
%

\subsection{Weak interactions}

In equations (\ref{eigenenergies}) and (\ref{rotation}), 
the rotation matrix was chosen such that the particle-hole excitation energy 
is always positive, $E^+_\mathbf{k}-E^-_\mathbf{k}>0$.
This choice is useful in the limit of strong interactions, see below.
However, in the limit of weak interactions, $U/J\ll1$, 
the Hubbard bands are overlapping and the system is in a metallic state where the notion of 
quasi-particles and holes looses its meaning.
In the weak-coupling limit, the calculation is simplified considerably 
if the rotation matrix orders the eigenvalues such that
\begin{align}\label{evweak}
E^-_\mathbf{k}&\approx J_\mathbf{k}+\frac{U}{2}\\
E^+_\mathbf{k}&\approx \frac{U}{2}\,.
\end{align}
Note that we never used the explicit form of the rotation matrix in the 
above derivation of the Boltzmann equation, therefore 
we have some freedom as long as the eigenvalue equation (\ref{eigenvalue}) is satisfied.
Equation (\ref{evweak}) corresponds to the choice
\begin{align}
O^a_X(\mathbf{k})\approx\frac{1}{\sqrt{2}}
\begin{pmatrix}
1+\frac{U}{2 J_\mathbf{k}} & 1-\frac{U}{2 J_\mathbf{k}} \\
-1+\frac{U}{2 J_\mathbf{k}} & 1+\frac{U}{2 J_\mathbf{k}}
\end{pmatrix}\,.
\end{align}
For $U/J\ll 1$, the dominating channel is $a=b=c=d=-$. 
The remaining matrix elements determine the dynamics of slower 
collisions with energies $\sim U^2/J$ or $\sim U$.
Using the energy conserving delta-distribution for the dominating channel, we find
from (\ref{4pcorr})
\begin{align}
\mathcal{A}^{----}_{-\mathbf{k}-\mathbf{q}-\mathbf{p},\mathbf{p},\mathbf{q},\mathbf{k}, \bar{s}\bar{s}ss}
=- \frac{U}{4}\left[f^-_{\mathbf{p},\bar{s}} f^-_{\mathbf{k},s}\left(1-f^-_{\mathbf{k}+\mathbf{q}+\mathbf{p},\bar{s}}\right)
\left(1-f^-_{{\mathbf{q}},s}\right)-f^-_{\mathbf{k}+\mathbf{q}+\mathbf{p},\bar{s}}
f^-_{\mathbf{q},s}\left(1-f^-_{\mathbf{p},\bar{s}}\right)\left(1-f^-_{\mathbf{k},s}\right)
\right]\,.
\end{align}
The evolution equation (\ref{boltzmannevolution}) simplifies to
\begin{align}
\partial_t  f^-_{\mathbf{k},a}&=-\frac{2\pi U^2}{N^2}\sum_{\mathbf{q},\mathbf{p}}
\delta\left(J_{\mathbf{k}+\mathbf{q}+\mathbf{p}}-J_\mathbf{p}+J_\mathbf{q}-J_\mathbf{k}\right)\nonumber\\
&\times\left[f^-_{\mathbf{p},\bar{s}} f^-_{\mathbf{k},s}\left(1-f^-_{\mathbf{k}+\mathbf{q}+\mathbf{p},\bar{s}}\right)
\left(1-f^-_{{\mathbf{q}},s}\right)-f^-_{\mathbf{k}+\mathbf{q}+\mathbf{p},\bar{s}}
f^-_{\mathbf{q},s}\left(1-f^-_{\mathbf{p},\bar{s}}\right)\left(1-f^-_{\mathbf{k},s}\right)
\right]\,.
\end{align}
In this limit the distribution function reads
\begin{align}
f^-_{\mathbf{k},s}=\frac{1}{2}+f_{\mathbf{k},s}^{00,\mathrm{corr}}+f_{\mathbf{k},s}^{10,\mathrm{corr}}+f_{\mathbf{k},s}^{01,\mathrm{corr}}+f_{\mathbf{k},s}^{11,\mathrm{corr}}=n_{\mathbf{k},s}
\end{align}
and we find
\begin{align}
\partial_t  n_{\mathbf{k},s}&=-\frac{2\pi U^2}{N^2}\sum_{\mathbf{q},\mathbf{p}}
\delta\left(J_{\mathbf{k}+\mathbf{q}+\mathbf{p}}-J_\mathbf{p}+J_\mathbf{q}-J_\mathbf{k}\right)
\left[n_{\mathbf{p},\bar{s}} n_{\mathbf{k},s}\left(1-n_{\mathbf{k}+\mathbf{q}+\mathbf{p},\bar{s}}\right)
\left(1-n_{{\mathbf{q}},s}\right)-n_{\mathbf{k}+\mathbf{q}+\mathbf{p},\bar{s}}
n_{\mathbf{q},s}\left(1-n_{\mathbf{p},\bar{s}}\right)\left(1-n_{\mathbf{k},s}\right)
\right]
\end{align}
which is the standard expression of the Boltzmann kinetic equations in the weak coupling limit.
It coincides with the perturbative result (\ref{boltzmannweak}) 
for $V_\mathbf{q}^{ss}=0$ and $V_\mathbf{q}^{s\s{s}}=U$.

\subsection{Strong interactions}

In the limit of strong interactions $J/U\ll 1$, we choose the rotation matrix 
$O^{a}_{X}(\mathbf{k})$ such that $E^+_\mathbf{k}-E^-_\mathbf{k}>0$.
From (\ref{rotation}) we find then 
$O^a_X(\mathbf{k})\approx\delta^a_X$.
The four-point correlator (\ref{4pcorr}) simplifies to
\begin{align}
\mathcal{A}^{abcd}_{-\mathbf{k}-\mathbf{q}-\mathbf{p},\mathbf{p},\mathbf{q},\mathbf{k}, \bar{s}\bar{s}ss}
=&\frac{1}{16}\bigg\{J_{\mathbf{k}+\mathbf{q}+\mathbf{p}}\left[-(-1)^b \delta^{cd}+(-1)^c \delta^{b\bar{d}}-
(-1)^d \delta^{bc}\right]+J_\mathbf{p}\left[-(-1)^a \delta^{cd}-(-1)^c \delta^{ad}
+(-1)^d \delta^{a\bar{c}}\right]\nonumber\\
&+J_\mathbf{q}\left[(-1)^a \delta^{b\bar{d}}-(-1)^b \delta^{ad}-(-1)^d \delta^{ab}\right]
+J_\mathbf{k}\left[-(-1)^a \delta^{bc}+(-1)^b \delta^{a\bar{c}}-(-1)^c\delta^{ab}\right]\bigg\}\nonumber\\
&\times \left[f^b_{\mathbf{p},\bar{s}} f^d_{\mathbf{k},s}\left(1-f^a_{\mathbf{k}+\mathbf{q}+\mathbf{p},\bar{s}}\right)
\left(1-f^c_{{\mathbf{q}},s}\right)-f^a_{\mathbf{k}+\mathbf{q}+\mathbf{p},\bar{s}}
f^c_{\mathbf{q},s}\left(1-f^b_{\mathbf{p},\bar{s}}\right)\left(1-f^d_{\mathbf{k},s}\right)
\right]\,.
\end{align}
From (\ref{boltzmannevolution}) follows then the evolution equation 
of the hole modes
\begin{align}
\partial_tf^-_{\mathbf{k},s}&=-\frac{2\pi}{N^2}\sum_{\mathbf{q},\mathbf{p}}
\delta\left(J_{\mathbf{k}+\mathbf{q}+\mathbf{p}}-J_\mathbf{p}+J_\mathbf{q}-J_\mathbf{k}\right)\nonumber\\
&\times\bigg\{(J_\mathbf{q}+J_{\mathbf{k}+\mathbf{q}+\mathbf{p}})^2
\left[f^-_{\mathbf{k},s} f^-_{\mathbf{p},\bar{s}}\left(1-f^-_{\mathbf{k}+\mathbf{q}+\mathbf{p},s}\right)
\left(1-f^-_{{\mathbf{q}},\bar{s}}\right)-f^-_{\mathbf{k}+\mathbf{q}+\mathbf{p},s}
f^-_{\mathbf{q},\bar{s}}\left(1-f^-_{\mathbf{k},s}\right)\left(1-f^-_{\mathbf{p},\bar{s}}\right)
\right]\nonumber\\
&+(J_\mathbf{q}-J_{\mathbf{p}})^2\left[f^-_{\mathbf{k},s} f^+_{\mathbf{p},\bar{s}}\left(1-f^+_{\mathbf{k}+\mathbf{q}+\mathbf{p},s}\right)
\left(1-f^-_{{\mathbf{q}},\bar{s}}\right)-f^+_{\mathbf{k}+\mathbf{q}+\mathbf{p},s}
f^-_{\mathbf{q},\bar{s}}\left(1-f^-_{\mathbf{k},s}\right)\left(1-f^+_{\mathbf{p},\bar{s}}\right)
\right]\nonumber\\
&+(J_{\mathbf{k}+\mathbf{q}+\mathbf{p}}-J_{\mathbf{p}})^2\left[f^-_{\mathbf{k},s} f^+_{\mathbf{p},\bar{s}}\left(1-
f^-_{\mathbf{k}+\mathbf{q}+\mathbf{p},s}\right)
\left(1-f^+_{{\mathbf{q}},\bar{s}}\right)-f^-_{\mathbf{k}+\mathbf{q}+\mathbf{p},s}
f^+_{\mathbf{q},\bar{s}}\left(1-f^-_{\mathbf{k},s}\right)\left(1-f^+_{\mathbf{p},\bar{s}}\right)
\right]\bigg\}
\end{align}
and for the particle modes (cf.~equation (\ref{boltzmann}))
\begin{align}
\partial_tf^+_{\mathbf{k},s}&=-\frac{2\pi}{N^2}\sum_{\mathbf{q},\mathbf{p}}
\delta\left(J_{\mathbf{k}+\mathbf{q}+\mathbf{p}}-J_\mathbf{p}+J_\mathbf{q}-J_\mathbf{k}\right)\nonumber\\
&\times\bigg\{(J_\mathbf{q}+J_{\mathbf{k}+\mathbf{q}+\mathbf{p}})^2
\left[f^+_{\mathbf{k},s} f^+_{\mathbf{p},\bar{s}}\left(1-f^+_{\mathbf{k}+\mathbf{q}+\mathbf{p},s}\right)
\left(1-f^+_{{\mathbf{q}},\bar{s}}\right)-f^+_{\mathbf{k}+\mathbf{q}+\mathbf{p},s}
f^+_{\mathbf{q},\bar{s}}\left(1-f^+_{\mathbf{k},s}\right)\left(1-f^+_{\mathbf{p},\bar{s}}\right)
\right]\nonumber\\
&+(J_\mathbf{q}-J_{\mathbf{p}})^2\left[f^+_{\mathbf{k},s} f^-_{\mathbf{p},\bar{s}}\left(1-f^-_{\mathbf{k}+\mathbf{q}+\mathbf{p},s}\right)
\left(1-f^+_{{\mathbf{q}},\bar{s}}\right)-f^-_{\mathbf{k}+\mathbf{q}+\mathbf{p},s}
f^+_{\mathbf{q},\bar{s}}\left(1-f^+_{\mathbf{k},s}\right)\left(1-f^-_{\mathbf{p},\bar{s}}\right)
\right]\nonumber\\
&+(J_{\mathbf{k}+\mathbf{q}+\mathbf{p}}-J_{\mathbf{p}})^2\left[f^+_{\mathbf{k},s} f^-_{\mathbf{p},\bar{s}}\left(1-
f^+_{\mathbf{k}+\mathbf{q}+\mathbf{p},s}\right)
\left(1-f^-_{{\mathbf{q}},\bar{s}}\right)-f^+_{\mathbf{k}+\mathbf{q}+\mathbf{p},s}
f^-_{\mathbf{q},\bar{s}}\left(1-f^+_{\mathbf{k},s}\right)\left(1-f^-_{\mathbf{p},\bar{s}}\right)
\right]\bigg\}\,.
\end{align}
Note that in the strong-coupling limit, the quasi-particle and hole distribution functions 
are related to the correlation functions via
\begin{align}
f^-_{\mathbf{k},s}&=1+2f^{00,\mathrm{corr}}_{\mathbf{k},s}\\
f^+_{\mathbf{k},s}&=2f^{11,\mathrm{corr}}_{\mathbf{k},s}\,.
\end{align}
\end{widetext}


\begin{thebibliography}{99}
\bibitem{footnote}
Here, we are mostly interested in the relaxation dynamics of the quasi-particles 
and holes (i.e., the charge modes) rather than the magnetic degrees of freedom 
(the spin modes). 
%
Thus, we consider a spin unpolarized state (without anti-ferrormagnetic ordering
as in the Mott-Neel state). 
%
Such a state can arise due to frustration (which prevents the anti-ferrormagnetic ordering)
or if the temperature is large enough to destroy spin ordering.
%
Note that the typical energy of the spin modes $\sim J^2/U$ is much smaller than the 
Mott gap $\sim U$ of the charge modes for large $U$, such that this temperature 
could still be relatively cold for the latter.
\end{thebibliography}
\end{document}